\documentclass[preprint]{aastex}
\slugcomment{accepted for publication in Astrophysical Journal}
\shorttitle{Searching for additional heating}
\shortauthors{B. Otte et al.}

\begin{document}

\title{Searching for additional heating --- [\ion{O}{2}] emission in the diffuse
ionized gas of \objectname[]{NGC\,891}\footnote{Based on observations made with
the William Herschel Telescope operated on the island of La Palma by the Isaac
Newton Group in the Spanish Observatorio del Roque de los Muchachos of the
Instituto de Astrofisica de Canarias.}, \objectname[]{NGC\,4631} and
\objectname[]{NGC\,3079}}
\author{B. Otte\footnote{Visiting Astronomer, Kitt Peak National Observatory,
National Optical Astronomy Observatories, which is operated by the Association
of Universities for Research in Astronomy, Inc. (AURA) under cooperative
agreement with the National Science Foundation.}, R. J. Reynolds, J. S.
Gallagher III\footnotemark[2]}
\affil{Department of Astronomy, University of Wisconsin--Madison}
\affil{475 North Charter Street, Madison, WI 53706}
\email{otte@astro.wisc.edu, reynolds@astro.wisc.edu, jsg@astro.wisc.edu}
\and
\author{A. M. N. Ferguson}
\affil{Kapteyn Astronomical Institute, University of Groningen}
\affil{P.O. Box 800, 9700 AV Groningen, The Netherlands}
\email{ferguson@astro.rug.nl}

\begin{abstract}
We present spectroscopic data of ionized gas in the disk--halo regions of three
edge--on galaxies, \objectname[]{NGC\,891}, \objectname[]{NGC\,4631} and
\objectname[]{NGC\,3079}, covering a wavelength range from
[\ion{O}{2}]\,$\lambda$3727\AA\ to [\ion{S}{2}]\,$\lambda$6716.4\AA. The
inclusion of the [\ion{O}{2}] emission provides new constraints on the
properties of the diffuse ionized gas (DIG), in particular, the origin of the
observed spatial variations in the line intensity ratios. We used three
different methods to derive electron temperatures, abundances and ionization
fractions along the slit. The increase in the [\ion{O}{2}]/H$\alpha$ line ratio
towards the halo in all three galaxies requires an increase either in electron
temperature or in oxygen abundance. Keeping the oxygen abundance constant yields
the most reasonable results for temperature, abundances, and ionization
fractions. Since a constant oxygen abundance seems to require an increase in
temperature towards the halo, we conclude that gradients in the electron
temperature play a significant role in the observed variations in the optical
line ratios from extraplanar DIG in these three spiral galaxies.
\end{abstract}


\section{INTRODUCTION}

When examining ionized gas, it is common practice to distinguish between
classical \ion{H}{2} regions (Str\"omgren spheres around OB stars) and diffuse
ionized gas (DIG), the gas outside the boundaries of the Str\"omgren spheres.
While \ion{H}{2} regions are created by photoionization, the ionization
processes for the DIG are less well known. Many attempts have been made to
explain the DIG by photoionization models (e.g. Domg\"orgen \& Mathis 1994;
Sokolowski 1992), while a few studies address the possibility of shock
excitation (e.g. Sivan, Stasi\'nska, \& Lequeux 1986). In recent years, DIG was
found not only in the disks of galaxies, but also far above the stellar disks in
the halo of the \objectname[]{Milky Way} and in several edge--on galaxies at
heights of more than 1\,kpc (e.g. Reynolds 1985; Rand, Kulkarni, \& Hester
1990). Questions therefore arise about where this extraplanar DIG (eDIG) comes
from and how it is ionized.

Dynamical models of galaxies like `galactic fountains' \citep{shap} and
`chimneys' \citep{norman} describe how gas can be transported from the disk into
the halo. Supernova explosions heating the gas in the disk and pushing it up
into the halo are important for both the dynamics and the ionization of the gas
in the halo. Due to the high velocities in this ejected gas, shocks can arise
and ionize the gas far above the disk. The models of runaway O stars leaving the
disk and moving into the halo (e.g. Gies 1987), leaking ionizing photons from
the disk into the halo due to low density gas (Miller \& Cox 1993; Dove \& Shull
1994) as well as the theory of photons created by neutrino decay \citep{sciama}
are further attempts to explain the ionization of extraplanar DIG.

Both \citet{martin} and \citet{rand} were able to explain the run of several of
the observed emission line ratios from \ion{H}{2} regions to the DIG with
composite models. Their models consisted of photoionization and one additional
ionization process (shock ionization, turbulent mixing layers). This additional
process was needed to explain the rise in the [\ion{O}{3}]/H$\beta$ line ratio
with increasing distance from the disk. However, even with these composite
models it was not possible to explain the constant [\ion{S}{2}]/[\ion{N}{2}]
line ratio, which was observed in \objectname[]{NGC\,891} \citep{rand}, as well
as in the \objectname[]{Milky Way} \citep{haffner} and other galaxies
\citep{otte}. These data led \citet{haffner} to the conclusion that the electron
temperature increases with increasing distance from the midplane of the
galaxies.

A rise in temperature can explain both the growing [\ion{O}{3}]/H$\beta$ ratio
as well as the constant [\ion{S}{2}]/[\ion{N}{2}] ratio with increasing galactic
altitude $|z|$ without invoking an additional ionization mechanism at high
$|z|$. Such a rise in electron temperature also should effect the
[\ion{O}{2}]/H$\alpha$ line ratio. The [\ion{O}{2}]\,$\lambda$3727\AA\ emission
line provides important additional information about the ionization and heating
processes in the DIG because of its high excitation energy. Below we present the
results of observations of [\ion{O}{2}], [\ion{O}{3}], H$\beta$, [\ion{N}{2}],
H$\alpha$, and [\ion{S}{2}] emission from the eDIG of three edge--on galaxies,
\objectname[]{NGC\,891}, \objectname[]{NGC\,4631} and \objectname[]{NGC\,3079}.
These objects represent the first targets of a small sample of edge--on galaxies
which have been chosen for their known eDIG emission. The analysis of the other
galaxies in our sample is still in progress. The results obtained from
\objectname[]{NGC\,891}, \objectname[]{NGC\,4631} and \objectname[]{NGC\,3079}
provide some evidence for an increase in temperature with increasing height.
Additional information is obtained about variations in the ionization state and
chemical abundances within the gaseous halos.

\section{OBSERVATIONS AND DATA REDUCTION}

The spectra of \objectname[]{NGC\,891} were obtained with the ISIS spectrograph
at the William Herschel 4.2\,m Telescope on La Palma, Canary Islands, in 1999
September 10 under photometric conditions and dark skies. Gratings R316R and
R300B were used for the red and the blue arm, respectively. Blocking filter
GG495 was used for the red arm. The slitwidth was 1''. The blue arm was read out
in a 2x2 binned mode and yielded pixel scales of 0$\farcs$40/pixel or
1.74\AA/pixel. The covered wavelength range was from about 3600\,\AA\ to
5400\,\AA. The red arm was rebinned after reduction and calibration to match
the spatial pixel scale of the blue arm. The wavelength dispersion for the red
arm was 1.47\AA/pixel with a wavelength range from about 5700\,\AA\ to
7200\,\AA. We combined two 30\,min and one 20\,min exposures in each arm. The
slit position is the same as in \citet{rand} and shown in Figure 1.

The spectra of \objectname[]{NGC\,4631} and \objectname[]{NGC\,3079} were
obtained with the GoldCam spectrograph at the 2.1\,m telescope on Kitt Peak, AZ,
in 2000 February 29 -- March 6. We used grating 9 with decker 4 and a slitwidth
of 2$\farcs$5. This yielded a pixel scale of 0$\farcs$80/pixel or
2.44\,\AA/pixel, respectively, and a wavelength range from about 3500\,\AA\ to
7400\,\AA. We used filter WG345 to remove possible overlaps between orders.
Exposure times varied from 20\,min to 30\,min depending on the weather. For
\objectname[]{NGC\,4631}, we combined eleven spectra with a total integration
time of 4.5 hours. For \objectname[]{NGC\,3079}, we combined nine spectra with a
total integration time of 3 hours and 20 minutes. Both slit positions are
perpendicular to the plane of the galaxies and shown in Figures 2 and 3. During
the observations of galaxies, we also took a few spectra of blank sky regions.

The spectra were reduced using standard procedures in IRAF\footnote{IRAF is
distributed by the National Optical Astronomy Observatories, which are operated
by the Association of Universities for Research in Astronomy, Inc., under
cooperative agreement with the National Science Foundation.}. For sky
subtraction, we subtracted differently scaled sky spectra from the red and the
blue part of the galaxy spectra of \objectname[]{NGC\,4631} and
\objectname[]{NGC\,3079}. This was necessary for the spectra taken during the
first night, which was partly cloudy. The sky spectra were smoothed along the
slit by 15 pixels to increase the per--pixel signal--to--noise ratios. The
resulting spectra are of very good quality, and, in particular, we appear to
have successfully removed the sky lines that can be confused with [\ion{O}{2}]
emission. For sky subtraction in our \objectname[]{NGC\,891} spectra, we
averaged over about 100 rows in the galaxy spectra which did not show any
galactic emission and used this average spectrum as sky. The spectra were then
calibrated in wavelength and corrected for distortion using standard star
exposures at different positions along the slit. After the flux calibration, we
combined the spectra at each slit position by carefully examining the positions
of emission lines in the wavelength direction and the position of flux features
in the spatial direction. Individual spectra were shifted by integer pixels, if
necessary, to overlap features and thus avoid broadening of lines and features
during the combining procedure. The routine for cosmic ray removal during
combining (averaging) of the spectra was insufficient. We therefore did not
remove cosmic rays during image combining, but cleared the areas around emission
lines of cosmic rays by hand later, before measuring emission lines. We only
used galaxy spectra of comparable signal--to--noise for the combining procedure,
that is we did not include spectra whose count rates were significantly reduced
due to obscuration by clouds. Fortunately, for most spectra, the periods of
cloud obscuration were short in comparison with the long exposure times. Thus,
variations in the count rates of the galaxy emission were small and therefore
negligible for most spectra. However, the variations in sky emission due to the
clouds during our 2.1\,m run could not be neglected, which we considered in our
sky subtraction described above. 

The analysis was done in MIDAS\footnote{MIDAS is developed and maintained by the
European Southern Observatory.}. To measure emission lines, we fitted Gaussian
curves to the emission lines at each row along the slit. We averaged each
emission line along the slit, if necessary, to obtain a better signal--to--noise
ratio in the halo. Usually, the measurements up to about 1.2\,kpc above the disk
were not averaged. For measurements between about 1.2\,kpc and 2\,kpc, we
averaged over five rows in the spatial direction, which corresponds to 75\,pc
in \objectname[]{NGC\,891}, 110\,pc in \objectname[]{NGC\,4631}, and 270\,pc in
 \objectname[]{NGC\,3079}. Above 2\,kpc, we usually averaged over nine rows
corresponding to 150\,pc, 215\,pc, and 540\,pc, respectively. The averaging was
done by comparing unaveraged measurements with averaged measurements to avoid
smoothing out features in the emission line fluxes, i.e. our averaging should
not have any effect on the derived line ratios.

The Balmer absorption lines could clearly be seen in the stellar continua around
the Balmer emission lines H$\beta$ and H$\gamma$ (see Fig. 4). To correct for
the underlying Balmer absorption lines, we assumed a constant equivalent width
for these absorption lines and an intrinsic emssion line intensity ratio of
H$\gamma$/H$\beta=0.466$. This procedure has been used, e.g., by \citet{liu}. We
integrated the flux over a wavelength region which included the absorption line
wings and could then determine the equivalent width with the assumptions
mentioned above. The derived equivalent width was $(4\pm1)$\,\AA\ along the
slit. The color of the galaxy continua did not change significantly with height
above the disk, suggesting that even at larger distances from the midplane,
where no reliable H$\gamma$/H$\beta$ ratios could be calculated, the equivalent
width of the Balmer absorption lines did not change signifiantly either. All
H$\beta$ absorption lines seemed to be $75\pm5$\,\AA\ wide at the continuum
level. We therefore fitted the H$\beta$ absorption line using the derived
equivalent width and the base width of 75\,\AA\ to be able to correct our
H$\beta$ emission line fluxes for the underlying absorption. We did not try to
correct the H$\alpha$ emission for H$\alpha$ absorption, because due to the low
resolution, the H$\alpha$ absorption line could not be observed, and was
probably partially filled by [\ion{N}{2}] emission. In all three galaxies, the
H$\alpha$/H$\beta$ line ratio decreases below the theoretical value of about 2.9
farther up in the halo at about 2\,kpc. Since the H$\beta$ emission becomes very
faint at these heights, our absorption line correction seems to fail for very
weak H$\beta$ emission. We will therefore exclude data connected to
H$\alpha$/H$\beta<2.9$ in our analysis.

The high values of H$\alpha$/H$\beta$ in the midplane of the galaxies
(H$\alpha$/H$\beta\approx 8$ for \objectname[]{NGC\,891} and
\objectname[]{NGC\,4631} and H$\alpha$/H$\beta\approx 7$ for
\objectname[]{NGC\,3079}) also show that extinction correction in the disk was
necessary. We derived the optical depth $\tau$ from the H$\alpha$/H$\beta$ line
ratio and used the ``usual'' method for the extinction coefficients described in
\citet{mathis}. Figure 5 shows the line ratio [\ion{O}{2}]/H$\alpha$ for
\objectname[]{NGC\,4631} without extinction correction (upper panel) and after
extinction correction (lower panel). Due to the uncertainty of the extinction
correction, it is not clear how physical the increase of the [\ion{O}{2}] line
ratio caused by the extinction correction really is. We did not estimate any
error for the extinction correction, i.e. the error bars of the line ratios only
reflect the errors of our flux measurements. The maximum extinction correction
yielded a factor of 9--12 for [\ion{O}{2}] (!), a factor of 2.6--3.0 for
H$\beta$, and a factor of 2.4--2.7 for [\ion{O}{3}], whereas the red lines had
factors of about 1. Since we were interested in the change of the line ratios
and the physical properties with increasing distance from the midplane, it was
important to keep in mind which values were affected the most and in what way by
the extinction correction.

\section{RESULTS}

For all three galaxies (\objectname[]{NGC\,891}, \objectname[]{NGC\,4631},
\objectname[]{NGC\,3079}), we wanted to fit the line ratios
[\ion{O}{2}]/H$\alpha$, [\ion{O}{3}]/H$\alpha$, [\ion{N}{2}]/H$\alpha$ and
[\ion{S}{2}]/H$\alpha$ from the eDIG. Since each line ratio depends on element
abundance, ionization fraction and temperature, we had to make a few assumptions
to decrease the number of unknowns: 1) We assumed that all the emitting gas is
ionized. This means in particular that H$^+/{\rm H}=1$. 2) Since the ionization
potentials of neutral nitrogen and oxygen are similar, we assumed that the
ionization fractions of both elements are the same for singly ionized atoms,
i.e. N$^+/{\rm N}={\rm O}^+$/O (e.g. Sembach et al. 2000). 3) It has been
observed in several objects that the ratio of the oxygen abundance to the sulfur
abundance is about 100/3, independent of metallicity \citep{allen}. We adopted
this ratio to derive the sulfur abundance from the oxygen abundance. 4) We
assumed that O$^{+++}$/O (and higher ionization stages of oxygen) is negligibly
small and therefore (O$^+/{\rm O})+({\rm O}^{++}/{\rm O})=1$. (No assumptions
were made for doubly ionized sulfur or doubly ionized nitrogen, i.e. we did not
derive nor use these ionization fractions.) 5) One additional assumption was
necessary, which yielded three different fitting methods. We adopted a constant
ionization fraction O$^{++}$/O along the slit (Method A), a constant temperature
along the slit (Method B), and a constant oxygen abundance along the slit
(Method C). In the following, we will discuss each method and its results in
more detail.

\subsection{Method A: O$^{++}$/O=const.}

For each fitting method, we started with the [\ion{O}{2}]/[\ion{O}{3}] line
ratio, because it depends only on temperature and ionization fractions, but not
on abundance. Due to the low resolution in our spectra, we were not able to
distinguish between [\ion{O}{2}]\,$\lambda$3726.0\AA\ and
[\ion{O}{2}]\,$\lambda$3728.8\AA. Thus, we use [\ion{O}{2}]\,$\lambda$3727\AA\
to mean the sum of both [\ion{O}{2}] doublet lines. The
[\ion{O}{2}]/[\ion{O}{3}] line ratio can be written as
\begin{equation}
\frac{I_{3727}}{I_{5006.9}} = 2.49\,{\rm e}^{-0.99/T_4}\,\left(\frac{O^+}{O}
\right)\,\left(\frac{O^{++}}{O}\right)^{-1}
\end{equation}
\citep{oster}. All line ratio equations contain intensities in
ergs\,s$^{-1}$\,cm$^{-2}$\,sr$^{-1}$ with $T_4$ being the electron temperature
in 10\,000\,K.

We tried four different ionization fractions for O$^{++}$/O (0.05, 0.10, 0.15
and 0.20) to cover a reasonable range and calculated the electron temperature
for each case from equation (1). We then used the [\ion{O}{3}]/H$\beta$ line
ratio
\begin{equation}
\frac{I_{5006.9}}{I_{4861.3}} = 5.03\cdot10^5\,T_4^{0.33}\,{\rm e}^{-2.88/T_4}
\,\left(\frac{O^{++}}{O}\right)\,\left(\frac{O^{\rule{0cm}{1ex}}}{H}\right)\,
\left(\frac{H^+}{H}\right)^{-1}
\end{equation}
\citep{oster} to derive the oxygen abundance O/H for each case. With the
assumption of similar ionization fractions for nitrogen and oxygen as mentioned
earlier, we could derive the nitrogen abundances from the
[\ion{O}{2}]/[\ion{N}{2}] line ratio
\begin{equation}
\frac{I_{3727}}{I_{6583.4}} = 2.65\,{\rm e}^{-1.69/T_4}\,\left(\frac{O^+}{O}
\right)\,\left(\frac{O^{\rule{0cm}{1ex}}}{H}\right)\,\left(\frac{N^+}{N}\right)
^{-1}\,\left(\frac{N^{\rule{0cm}{1ex}}}{H}\right)^{-1}
\end{equation}
\citep{oster}. We also could derive the ionization fractions of singly ionized
sulfur for each case by assuming the ratio O/S to be 100/3 (as mentioned above)
and using the [\ion{O}{2}]/[\ion{S}{2}] line ratio
\begin{equation}
\frac{I_{3727}}{I_{6716.4}} = 0.58\,{\rm e}^{-1.73/T_4}\,\left(\frac{O^+}{O}
\right)\,\left(\frac{O^{\rule{0cm}{1ex}}}{H}\right)\,\left(\frac{S^+}{S}\right)
^{-1}\,\left(\frac{S^{\rule{0cm}{1ex}}}{H}\right)^{-1}
\end{equation}
\citep{oster}.
The final step now was to compare the observed H$\alpha$ line ratios with the
predicted ratios based upon the derivations above:
\begin{eqnarray}
{\rm [O\,II]/H\alpha:}\quad\frac{I_{3727}}{I_{6562.8}} & = & 4.31\cdot10^5\,T_4
^{0.4}\,{\rm e}^{-3.87/T_4}\,\left(\frac{O^+}{O}\right)\,\left(\frac{O^{\rule
{0cm}{1ex}}}{H}\right)\,\left(\frac{H^+}{H}\right)^{-1} \\
{\rm [O\,III]/H\alpha:}\quad\frac{I_{5006.9}}{I_{6562.8}} & = & 1.74\cdot10^5\,
T_4^{0.4}\,{\rm e}^{-2.88/T_4}\,\left(\frac{O^{++}}{O}\right)\,\left(\frac{O^
{\rule{0cm}{1ex}}}{H}\right)\,\left(\frac{H^+}{H}\right)^{-1} \\
{\rm [N\,II]/H\alpha:}\quad\frac{I_{6583.4}}{I_{6562.8}} & = & 1.62\cdot10^5\,
T_4^{0.4}\,{\rm e}^{-2.18/T_4}\,\left(\frac{N^+}{N}\right)\,\left(\frac{N^{\rule
{0cm}{1ex}}}{H}\right)\,\left(\frac{H^+}{H}\right)^{-1} \\
{\rm [S\,II]/H\alpha:}\quad\frac{I_{6716.4}}{I_{6562.8}} & = & 7.49\cdot10^5\,
T_4^{0.4}\,{\rm e}^{-2.14/T_4}\,\left(\frac{S^+}{S}\right)\,\left(\frac{S^{\rule
{0cm}{1ex}}}{H}\right)\,\left(\frac{H^+}{H}\right)^{-1}
\end{eqnarray}
(Osterbrock 1989; Haffner, Reynolds, \& Tufte 1999).

The derived values for electron temperature, oxygen and nitrogen abundances and
for the ionization fraction of sulfur, as well as the fitted line ratios, are
shown for \objectname[]{NGC\,891} in Figure 6, for \objectname[]{NGC\,4631} in
Figure 7 and for \objectname[]{NGC\,3079} in Figure 8. Since the [\ion{O}{3}]
and H$\beta$ emission lines were fainter than the other emission lines in the
spectra, the line ratio fits, which are based on these lines, usually do not
extend as far out into the halo as the observed line ratios. The predicted
H$\alpha$ line ratios using different ionization fractions for doubly ionized
oxygen do not differ from each other very much. However, the derived
temperatures, abundances and sulfur ionization fractions are quite sensitive to
the ionization fraction of oxygen. This is true for all three galaxies.

The line ratio predictions for \objectname[]{NGC\,891} match the observed line
ratios quite well. The peaks in the [\ion{O}{2}]/H$\alpha$ line ratio in the
midplane were enhanced by the uncertain internal extinction corrections. In
\objectname[]{NGC\,891}, the range in height from the midplane affected by
internal extinction is about $|z|\leq 1200$\,pc. Our [\ion{N}{2}]/H$\alpha$ line
ratios match those observed by \citet{col} at the same slit position. However,
our [\ion{S}{2}]/H$\alpha$ line ratios are higher below the disk (1.0 at
$z=-2$\,kpc as opposed to 0.6), but the same above the disk (0.6 at
$z=+2$\,kpc). Our [\ion{O}{3}]/H$\alpha$ measurements are slightly higher at
$|z|=2$\,kpc than the values measured by Collins \& Rand, but certainly
comparable within the error bars. However, Collins \& Rand did not measure as
high a peak in [\ion{O}{3}]/H$\alpha$ at $z\approx 200$\,pc as we did.

In \objectname[]{NGC\,4631}, the line ratio predictions also match the observed
line ratios. The peaks in [\ion{O}{2}]/H$\alpha$ at $z\approx -1200$\,pc,
$z\approx -600$\,pc and $z\approx -300$\,pc were all introduced by the
extinction correction. The line ratio predictions for \objectname[]{NGC\,3079}
seem to match the observed line ratios as well. However, in the range
$-2200\,{\rm pc}<z<-600$\,pc, the H$\beta$ emission line in the underlying
Balmer absorption line could not clearly be distinguished from the noise.
Therefore, it was not possible to predict line ratios in this part of the slit.
And the derived optical depth for the extinction correction and with it the
[\ion{O}{2}]/H$\alpha$ and [\ion{O}{3}]/H$\alpha$ line ratios are rather
questionable in this range. Unfortunately, the nitrogen line at 6583.4\,\AA\
had been hit by cosmic rays in the range $-3.4\,{\rm kpc}<z<-2.6\,{\rm kpc}$ and
$+1.4\,{\rm kpc}<z<+2.2\,{\rm kpc}$, which then contaminated the
[\ion{N}{2}]/H$\alpha$ line ratio in these areas. Also, the [\ion{O}{3}]
emission line at 5006.9\,\AA\ had been hit by a cosmic ray at
$z<-2.8\,{\rm kpc}$.

In all three galaxies, the derived abundances for oxygen and nitrogen are higher
than solar by up to four orders of magnitude for the ionization fraction
O$^{++}/{\rm O}=0.05$. These abundances seem unlikely, because it is difficult
to obtain higher abundances farther away from star forming regions, especially
in the halo. The ionization fraction O$^{++}/{\rm O}=0.15$ and 0.20 yielded
sulfur ionization fractions close to or greater than unity. This is unlikely (or
even impossible) given the low (23\,eV) ionization potential of S$^+$ and the
significant amount of S$^{++}$ predicted by photoionization models (e.g. Mathis
1986). The electron temperature seems to be constant in the halo in all three
galaxies with higher values in the area affected by the extinction correction.
The only oxygen ionization fraction with reasonable values for all derived
properties is O$^{++}/{\rm O}=0.10$, except for S$^+$/S in
\objectname[]{NGC\,891}. No reasonable solution seems to exist for
\objectname[]{NGC\,891} using Method A.

\subsection{Method B: T$_4$=const.}

For the next method, we assumed a constant electron temperature along the slit
and derived the ionization fraction of oxygen from equation (1). We tried four
different temperatures ($T_4=0.4$, 0.6, 0.8 and 1.0). For each temperature, we
derived the oxygen and nitrogen abundances and the sulfur ionization fraction
using equations (2)--(4). We then compared the observed line ratios relative to
H$\alpha$ with the ratios predicted by equations (5)--(8). The predictions fit
the observations as well as in Method A and therefore are not shown here.
However, the derived values for ionization fractions and abundances again vary
considerably from case to case as shown in Figures 9--11.

The lowest electron temperature ($T_4=0.4$) yielded nitrogen and oxygen
abundances one or two orders of magnitude above solar in all three galaxies.
Even $T_4=0.6$ produced oxygen abundances above solar, whereas the nitrogen
abundances have approximately solar values. The highest temperature ($T_4=1.0$)
caused very high or unreasonable sulfur ionization fractions in the halo of the
galaxies. S$^+$/S is still somewhat high for $T_4=0.8$ in
\objectname[]{NGC\,891}. Therefore, we find the most reasonable values for
abundances and ionization fractions for an electron temperature of
$T_4\approx0.7$. By keeping the electron temperature constant, the nitrogen
abundances increase towards the halos in all three galaxies, whereas the oxygen
abundance increases slightly or almost stays constant with increasing distance
from the midplane.

\subsection{Method C: O/H=const.}

Our last fitting method differs a little from the previous two methods. We
combined equations (1) and (2) to eliminate the dependence on the ionization
fractions of oxygen. [\ion{O}{3}]/H$\beta$ becomes
\begin{equation}
\frac{\rm [O\,III]}{\rm H\beta}=\frac{5.03\cdot 10^5\cdot T_4^{0.33}\cdot
{\rm e}^{-2.88/T_4}}{{\rm [O\,II]}/{\rm [O\,III]}\cdot 0.40\cdot{\rm e}^
{0.99/T_4}+1}\cdot\left(\frac{O^{\rule{0cm}{1ex}}}{H}\right).
\end{equation}
We assumed solar abundance for oxygen
(O/${\rm H}=8.5\cdot10^{-4}$ \citep{dapp}). We then calculated the minimum and
maximum allowed electron temperature for each data point along the slit by
fitting the observed [\ion{O}{3}]/H$\beta$ line ratios within their 1\,$\sigma$
error bars using temperature increments/decrements of 400\,K. For both
temperature extremes, we derived the nitrogen abundances and the oxygen and
sulfur ionization fractions using equations (2)--(4) and compared the predicted
with the observed H$\alpha$ line ratios as usual.

The results for the three galaxies are shown in Figures 12--14. We included
calculations for each galaxy which were not corrected for extinction to show how
sensitive the predicted line ratios as well as the derived properties are to
extinction corrections. For all three galaxies, the predicted line ratios using
the minimum electron temperature are generally too low in the halo in comparison
with the observed values. However, the maximum allowed temperature fits are
within the 1\,$\sigma$ error bars of the observed H$\alpha$ line ratios
([\ion{O}{3}]/H$\alpha$, [\ion{N}{2}]/H$\alpha$) or slightly above
([\ion{O}{2}]/H$\alpha$, [\ion{S}{2}]/H$\alpha$) in the halo. Subtracting about
a fourth of the difference between maximum and minimum temperature from the
maximum allowed temperature seems to fit the line ratios best. The only
exception is for the uncorrected line ratios below the disk of
\objectname[]{NGC\,4631} (down to $z=-1400$\,kpc), where the predicted line
ratios of both minimum and maximum electron temperature are too high for all
four line ratios. We point out that the previously derived optical depth is
greater than zero in this region. In general, the extinction corrected line
ratios and their predictions match better than the not extinction corrected
values in areas where the derived $\tau>0$. This is especially true for the
region around the midplane of each galaxy.

The derived nitrogen abundances are close to solar for both temperatures in all
three galaxies. The sulfur ionization fractions are below unity for both
temperature ranges in \objectname[]{NGC\,891}, \objectname[]{NGC\,4631} and
\objectname[]{NGC\,3079}, but exceed unity in the uncorrected case of
\objectname[]{NGC\,891}. The comparison between the extinction corrected values
and the uncorrected values for each galaxy shows clearly, how strongly the
extinction correction affects the derived properties, in particular the nitrogen
abundances and the sulfur ionization fractions.

In all three galaxies, the extinction corrected values of S$^+$/S increase with
$|z|$, where $\tau>0$, and then decrease or stay constant, where $\tau=0$. The
uncorrected values of S$^+$/S decrease with $|z|$ (\objectname[]{NGC\,891},
\objectname[]{NGC\,3079}) or stay constant (\objectname[]{NGC\,4631}). The same
behaviour can be seen in the nitrogen abundance of each galaxy, i.e. increase
with $|z|$, where $\tau>0$, and then decrease in N/H or constant N/H, where
$\tau=0$, and decrease with $|z|$ or constant N/H in the uncorrected cases. In
the extinction corrected calculations, O$^{++}$/O more or less increases with
distance from the midplane in all three galaxies. In the uncorrected cases,
O$^{++}$/O slightly increases with $|z|$ in \objectname[]{NGC\,891} and basicly
stays constant in \objectname[]{NGC\,4631} and \objectname[]{NGC\,3079}.

The electron temperature stays constant or increases with $|z|$ in all three
galaxies (not considering the peaks introduced by the extinction correction).
In \objectname[]{NGC\,891}, $T_4$ stays overall constant within the range
affected by the extinction correction ($|z|\leq 1200$\,pc) and then starts to
increase with $z$ at least above the disk ($z>+1200$\,pc). Below the disk, no
obvious increase is observed. In the uncorrected case of
\objectname[]{NGC\,891}, the electron temperature starts to increase already at
$|z|=400$\,pc from $T_4\approx 0.6$ to $T_4\approx 0.72$ at $z=+1600$\,pc and
from $T_4\approx 0.56$ to $T_4\approx 0.69$ at $z=-1.8$\,kpc (considering the
average of minimum and maximum allowed electron temperature).

In \objectname[]{NGC\,4631}, the extinction correction affects the range of
$-1400\,{\rm pc}<z<1000$\,pc. Beyond this range in height, the electron
temperature increases from $T_4\approx 0.7$ to $T_4\approx 0.8$ above the disk.
Below the disk, the electron temperature may slightly increase from
$T_4\approx 0.73$ to $T_4\approx 0.8$ at $z=-2$\,kpc. However, the temperature
may also be constant in this area given the range set by minimum and maximum
allowed temperature. In the uncorrected case of \objectname[]{NGC\,4631}, the
temperature increase is more obvious at least above the disk, ranging from
$T_4\approx 0.66$ at $z=+500$\,pc to $T_4\approx 0.8$ at $z=+2$\,kpc. Below the
disk, the temperature seems to increase from $T_4\approx 0.72$ at $z=-1$\,kpc to
$T_4\approx 0.8$ at $z=-2$\,kpc. However, the temperature may be constant in
this area considering the difference between minimum and maximum allowed
temperature.

The electron temperature seems to stay more or less constant below the disk of
\objectname[]{NGC\,3079} (ignoring the increase between $z=0$ and $z=-1$\,kpc
enhanced by the extinction correction). The temperature slightly increases with
$z$ above the disk (from $T_4=0.67$ at $z=0$ to $T_4=0.78$). In the uncorrected
case of \objectname[]{NGC\,3079}, the electron temperature increases from
$T_4=0.6$ at $z=0$ to $T_4\approx 0.78$ at $z=+2$\,kpc and $T_4\approx 0.7$ at
$z=-2.8$\,kpc. However, the variations (wiggles) in the temperatures in
\objectname[]{NGC\,3079} are larger relative to the increase in $T_4$ than in
the two other galaxies, making the temperature increases less convincing in
\objectname[]{NGC\,3079} than in \objectname[]{NGC\,891} and
\objectname[]{NGC\,4631}. We point out that there is no temperature decrease
with increasing distance from the midplane in any of the discussed cases.

The disadvantage of using the [\ion{O}{2}], [\ion{O}{3}] and H$\beta$ emission
lines for the derivation of $T_4$ is obviously the dependence on the extinction
correction. However, this dependence becomes less and even negligibly small in
the halo with increasing distance from the midplane. The advantage of using the
blue emission lines to derive the electron temperature is the stronger
dependence of the [\ion{O}{2}]/[\ion{O}{3}] and [\ion{O}{3}]/H$\beta$ line
ratios on $T_4$, i.e. these line ratios are more sensitive to the electron
temperature than the [\ion{N}{2}]/H$\alpha$ line ratio used by \citet{haffner}
or \citet{col}. Our (averaged) temperatures for \objectname[]{NGC\,891} are
1300\,K--1600\,K lower in the halo and 500\,K--900\,K lower in the disk than
those derived by Collins \& Rand (again ignoring the peaks introduced by our
extinction correction).

Since the assumption of solar oxygen abundance yielded already reasonable values
for temperature, nitrogen abundance and ionization fractions, we did not use
this method with other values for the oxygen abundance.

\section{DISCUSSION}

The overall trends of the H$\alpha$ line ratios in \objectname[]{NGC\,891},
\objectname[]{NGC\,4631} and \objectname[]{NGC\,3079} are the same as those in
other galaxies, i.e. [\ion{S}{2}]/H$\alpha$, [\ion{N}{2}]/H$\alpha$ and
[\ion{O}{3}]/H$\alpha$ increase with increasing height. [\ion{O}{2}]/H$\alpha$
also increases with distance from the midplane. In \objectname[]{NGC\,3079}, the
overall emission is fainter than in \objectname[]{NGC\,4631} and
\objectname[]{NGC\,891}.

The [\ion{S}{2}]/[\ion{N}{2}] line ratio in \objectname[]{NGC\,891} (Fig. 15,
top panel) stays constant from about $z=-1400$\,pc to about $z=+2600$\,pc at a
value of $0.55\pm0.10$. It matches the value measured by \citet{col}. The
[\ion{O}{2}]/[\ion{N}{2}] line ratio increases towards the halo in the
extinction corrected case (at $|z|>1$\,kpc, Fig. 15, middle panel) as well as
the uncorrected case (at $|z|>200-400$\,pc, Fig. 15, bottom panel). These
results are consistent with the variations in electron temperature in the
extraplanar diffuse ionized gas found by the analysis in section 3 above. These
temperature variations leave the [\ion{S}{2}]/[\ion{N}{2}] ratio in the eDIG
basicly unchanged, because this ratio is almost independent of electron
temperature, whereas the strongly electron temperature dependent ratio
[\ion{O}{2}]/[\ion{N}{2}] increases with $|z|$. Changes in the ionization
fractions may cause the smaller scale variations of [\ion{S}{2}]/[\ion{N}{2}]
and [\ion{O}{2}]/[\ion{N}{2}] along the slit.

In \objectname[]{NGC\,4631}, one unusual area is the \ion{H}{2} region at about
1\,kpc below the disk. Its [\ion{S}{2}]/H$\alpha$ and [\ion{N}{2}]/H$\alpha$
line ratios are lower than those of the surrounding diffuse gas (as expected by
photoionization), whereas the [\ion{O}{3}]/H$\alpha$ ratio only increases in one
part of the \ion{H}{2} region (at $z\approx-800$\,pc), but not in the other part
(at $z\approx-1$\,kpc). More observations are necessary to understand this
particular region. The line ratio [\ion{S}{2}]/[\ion{N}{2}] (Fig. 16, top panel)
seems to stay constant with increasing height in \objectname[]{NGC\,4631}
($0.8\pm0.1$ above $z=+1$\,kpc and $1.0\pm0.2$ below $z=+300$\,pc). In the
midplane, the value decreases to a minimum of 0.55 at about $z=+500$\,pc.
However, the strong variations in electron temperature (up to 5000\,K) below the
disk as derived in section 3 are big enough to enhance the variations below the
disk even in the almost temperature independent line ratio
[\ion{S}{2}]/[\ion{N}{2}].

The middle panel in Fig. 16 shows the extinction corrected
[\ion{O}{2}]/[\ion{N}{2}] line ratio in \objectname[]{NGC\,4631}, the bottom
panel shows the uncorrected line ratio. In the extinction corrected case, the
variations in the [\ion{O}{2}]/[\ion{N}{2}] line ratio below the disk are caused
by variations in electron temperature according to our analysis above (Method
C). In the uncorrected case, the [\ion{O}{2}]/[\ion{N}{2}] line ratio seems to
stay constant below the disk except for the unusual \ion{H}{2} region mentioned
above. The variations in electron temperature seem to be compensated by strong
variations in nitrogen abundance below the disk. Above the disk, the
[\ion{O}{2}]/[\ion{N}{2}] line ratio first stays constant
($+1000\,{\rm pc}<z<+1600$\,pc) and then increases with $z$ up to $z=+2$\,kpc
(in both the extinction corrected and the uncorrected case). According to our
results from Method C, this behaviour can be explained by a steady increase in
electron temperature combined with an increase in nitrogen abundance up to
$z=+1600$\,pc and a decrease in N/H between 1600\,pc and 2000\,pc.
Unfortunately, since our derived values for H$\alpha$/H$\beta$ drop below 2.9 at
$z=+2$\,kpc, we do not have reliable estimates for abundances or ionization
fractions in this area and therefore cannot explain the decrease in
[\ion{O}{2}]/[\ion{N}{2}] at $z>+2$\,kpc. Due to a superbubble above the disk of
\objectname[]{NGC\,4631} (Hoopes, Walterbros, \& Rand 1999), the conditions in
the eDIG might be different from those below the disk, resulting in the lower,
but still constant [\ion{S}{2}]/[\ion{N}{2}] ratio (in comparison with the area
below the disk) and the slower increase of [\ion{O}{2}]/[\ion{N}{2}] towards the
halo (in comparison with the observed increase in \objectname[]{NGC\,891}),
assuming that the temperature still dominates over photoionization in the
superbubble region.

The [\ion{S}{2}]/[\ion{N}{2}] ratio in \objectname[]{NGC\,3079} (Fig. 17, top
panel) stays constant with increasing height above the disk ($5.3\pm0.7$ at
$z>+200$\,pc), except for the area affected by the cosmic ray. However, below
the disk, the ratio first increases with height (from 0.45 at $z=0$ to about 1.2
at $z=-1.5$\,kpc), then decreases over a short distance to about 0.6 at
$z=-1.8$\,kpc. Beyond that point, [\ion{S}{2}]/[\ion{N}{2}] seems to increase
again with distance from the midplane. Unfortunately, we could not determine
abundances or ionization fractions between $-2$\,kpc and $-800$\,pc because of
too faint H$\beta$ emission.

The middle panel in Fig. 17 shows the extinction corrected
[\ion{O}{2}]/[\ion{N}{2}] line ratio in \objectname[]{NGC\,3079} along the slit,
the bottom panel again shows the uncorrected line ratio. In both cases, the
line ratio increases with distance from the midplane below the disk
($z<-1800$\,pc). Above the disk, [\ion{O}{2}]/[\ion{N}{2}] seems to stay
constant or to slightly decrease with $z$ at $z>+600$\,pc. According to our
Method C, the decrease in [\ion{O}{2}]/[\ion{N}{2}] between $z=0$ and
$z=+1400$\,pc in the extinction corrected case is caused by an increase in
nitrogen abundance (about a factor of 2), while the electron temperature
increases only slighty. In the uncorrected case, [\ion{O}{2}]/[\ion{N}{2}] first
increases between $z=0$ and $z=+600$\,pc, caused by an increase in electron
temperature combined with a decrease in nitrogen abundance, and then stays more
or less constant up to $z=+1400$\,pc. Neither electron temperature nor nitrogen
abundance vary significantly between $z=+600$\,pc and $z=+1400$\,pc.

In all three galaxies, the behaviour of [\ion{S}{2}]/[\ion{N}{2}] and
[\ion{O}{2}]/[\ion{N}{2}] can also be explained by variations of several orders
of magnitude in the abundances (according to our results of Method A) or
variations in the abundances combined with strong variations in the ionization
fractions (according to our results of Method B).

The question now is which of the three fitting methods yields the most
reasonable results. The best predictions in Method A (O$^{++}/{\rm O}={\rm
const.}$) are produced, if the ionization fraction O$^{++}/{\rm O}=0.10$. This
is true for two of our three galaxies. However, even with this value, the
derived abundances are rather unlikely, because they vary by more than one order
of magnitude along the slit. Moreover, it is difficult, if not impossible to
obtain oxygen and nitrogen abundances below or equal to solar along the entire
slit without exceeding unity for S$^+$/S.

In Method B ($T_4={\rm const.}$), the most reasonable value for $T$ appears to
be 7000\,K. However, the requirement of a non--varying temperature causes an
increase in abundance away from the midplane, although the resulting spread in
oxygen and nitrogen abundances are less than for Method A. The values for N/H
are below or close to solar in all three galaxies along the slit for electron
temperatures of about 7000\,K. The oxygen abundance is about solar along the
slit (slightly higher for \objectname[]{NGC\,4631}), but probably reaches higher
than solar values in the halo according to the slight increase of O/H with
$|z|$. The abundances increase with increasing distance from the disk. To
achieve this abundance gradient, outflows or chimneys with stable chimney walls
are needed. In this scenario, the metal rich gas from the disk is blown into the
halo to heights where the density of the ambient medium is low enough that
mixing can occur, while the higher density and higher pressure along the sides
of the outflows prevent the mixing of gas. However, it is rather unlikely that
only powerful outflows occur which can reach far up into the halo and enrich the
halo gas, while the diffuse gas near the disk retains lower metal abundances. It
is more likely that there are many more smaller scale outflows than large scale
ones, so that the halo gas near the disk becomes more metal enriched than the
halo gas farther away from the disk.

The nitrogen abundances in Method C (O/H\,=\,const.) are about solar or below
along the slit in \objectname[]{NGC\,4631} and \objectname[]{NGC\,3079}. In
\objectname[]{NGC\,891}, N/H reaches twice the solar value in the extinction
corrected case and higher values in the uncorrected case. The variations in N/H
are smaller than the corresponding values of Method A and B. N/H
stays constant or decreases with increasing distance from the midplane (ignoring
the overcorrected values of the extinction correction). The gradient is
consistent with the hypothesis of stars in the disk enriching the gas around
them and the gravitational potential of the disk keeping the enriched gas near
the disk. An almost constant abundance would be a sign of well--mixed gas.

The behaviour of the sulfur ionization fraction S$^+$/S depends strongly on the
extinction correction. S$^+$/S seems to stay constant (\objectname[]{NGC\,4631},
\objectname[]{NGC\,3079}) or to decrease (\objectname[]{NGC\,891}) with
increasing distance from the midplane, where $\tau=0$. A decrease in S$^+$/S
with $|z|$ would indicate an increase in S$^{++}$/S and thus an additional
source of ionization for the halo gas. The ionization potential for O$^{++}$ is
higher than that for S$^{++}$ (35\,eV and 23\,eV, respectively). Therefore, the
ionization fraction of doubly ionized sulfur should be at least as high as that
of doubly ionized oxygen, and thus (S$^+/{\rm S})+({\rm O}^{++}/{\rm O})\leqq1$
along the slit. This seems to be the case in all three galaxies for $T_4=0.7$ in
Method B and temperatures close to, but below the maximum allowed temperatures
along the slits in Method C. For the most reasonable ionization fraction
O$^{++}/{\rm O}=0.10$ in Method A, this condition is violated at a few data
points in \objectname[]{NGC\,4631} and \objectname[]{NGC\,3079} and almost
everywhere along the slit in \objectname[]{NGC\,891}. The maximum temperatures
reached in the halos of the galaxies in Method C are consistent with
temperatures observed in other galaxies and the \objectname[]{Milky Way}
($T_4\approx0.7-0.9$).

Photoionization models that use a decreasing ionization parameter with
increasing $|z|$ predict a decrease in O$^{++}$/O and an increase in S$^+$/S
with increasing distance from the midplane \citep{semb}. We do not observe a
decrease in electron temperature in any of our methods. An increasing electron
temperature towards the halo appears to require an additional heating mechanism.
If a harder source of ionization caused the observed constant or increasing
temperatures towards the halo, O$^{++}$/O would have to be constant or to
increase as well. We are not able to draw reliable conclusions about the
behaviour of the ionization fractions, as they can be completely alternated by
the applied extinction correction. However, the mostly constant
[\ion{S}{2}]/[\ion{N}{2}] line ratios we observed rule out any additional
heating mechanisms that also ionize the gas. Possible mechanisms which heat, but
do not ionize the gas are dissipation of turbulence, grain heating, Coulomb
collisions of cosmic rays, and magnetic field reconnections (Reynolds, Haffner,
\& Tufte 2000).

The eDIG in the three galaxies we observed behaves similarly within each
method, even though the derived values for abundances, ionization fractions and
temperatures differ slightly, and each galaxy also shows some peculiarities.
Given these observations and derivations, Method C yields the most reasonable
values, that is, a relatively uniform, near solar abundance distribution within
the galaxies (except \objectname[]{NGC\,891}). The extinction correction is
questionable, with indications that it overcorrects near the midplane in the
blue part of the spectrum; however, some correction is necessary given the high
H$\alpha$/H$\beta$ line ratios in the disk. The extinction also raises the
question whether the emission of all the spectral lines originated in the same
spatial region in the galaxies or whether for example the [\ion{O}{2}] photons
we observed were emitted from different gas clouds than the H$\beta$ photons.
Thus, the true values for the emission line ratios and the derived properties
near the midplane may differ from those shown here, but we expect the errors
will not be large ones at least in the halo. The key questions therefore are how
high the midplane peaks in the [\ion{O}{2}]/H$\alpha$ line ratios really become
and what causes them. Faint [\ion{O}{1}] emission also appears to be present in
the eDIG of \objectname[]{NGC\,4631} and \objectname[]{NGC\,3079}, suggesting
that neutral gas is present beyond the disks of these galaxies. Unfortunately,
this emission is too faint to measure from our data.

\section{CONCLUSIONS}

We explored three different methods to explain the observed line ratio
variations in the three galaxies, \objectname[]{NGC\,891},
\objectname[]{NGC\,4631} and \objectname[]{NGC\,3079}. Overall, one can say that
the line ratios require either a constant or increasing electron temperature
within the eDIG towards the halos, or an increase in abundances. Assuming a
constant (near solar) oxygen abundance along the slit (our Method C, section
3.3) yields the most reasonable values for ionization fractions, abundances and
temperature of these three methods. The derived values contradict pure
photoionization models. We conclude therefore that variations in the electron
temperature play a significant role in observed variations in the optical line
intensity ratios from the eDIG in these three galaxies.

\acknowledgements

B. O. is thankful to the staff on Kitt Peak for their support during the
observations. She is also grateful to S. Jansen for his computer support. The
authors thank R. J. Rand for information on his NGC\,891 data and the referee
R. A. M. Walterbos for his comments and suggestions resulting in an improved
paper. This reasearch was funded by the NSF through grant AST96--19424 and in
part by the Graduate School of the University of Wisconsin--Madison. The
observing run at the 2.1\,m was funded in part by the National Optical Astronomy
Observatory. The Digitized Sky Surveys were produced at the Space Telescope
Science Institute under U. S. Government grant NAGW-2166. The images of these
surveys are based on photographic data obtained using the Oschin Schmidt
Telescope on Palomar Mountain. The National Geographic Society -- Palomar
Observatory Sky Atlas was made by the California Institute of Technology with
grants from the National Geographic Society. The Second Palomar Observatory Sky
Survey was made by the California Institute of Technology with funds from the
National Science Foundation, the National Geographic Society, the Sloan
Foundation, the Samuel Oschin Foundation, and the Eastman Kodak Corporation. The
Oschin Schmidt Telescope is operated by the California Institute of Technology
and Palomar Observatory. Supplemental funding for sky--survey work at the STScI
is provided by the European Southern Observatory.

\pagebreak

\begin{figure}
\plotone{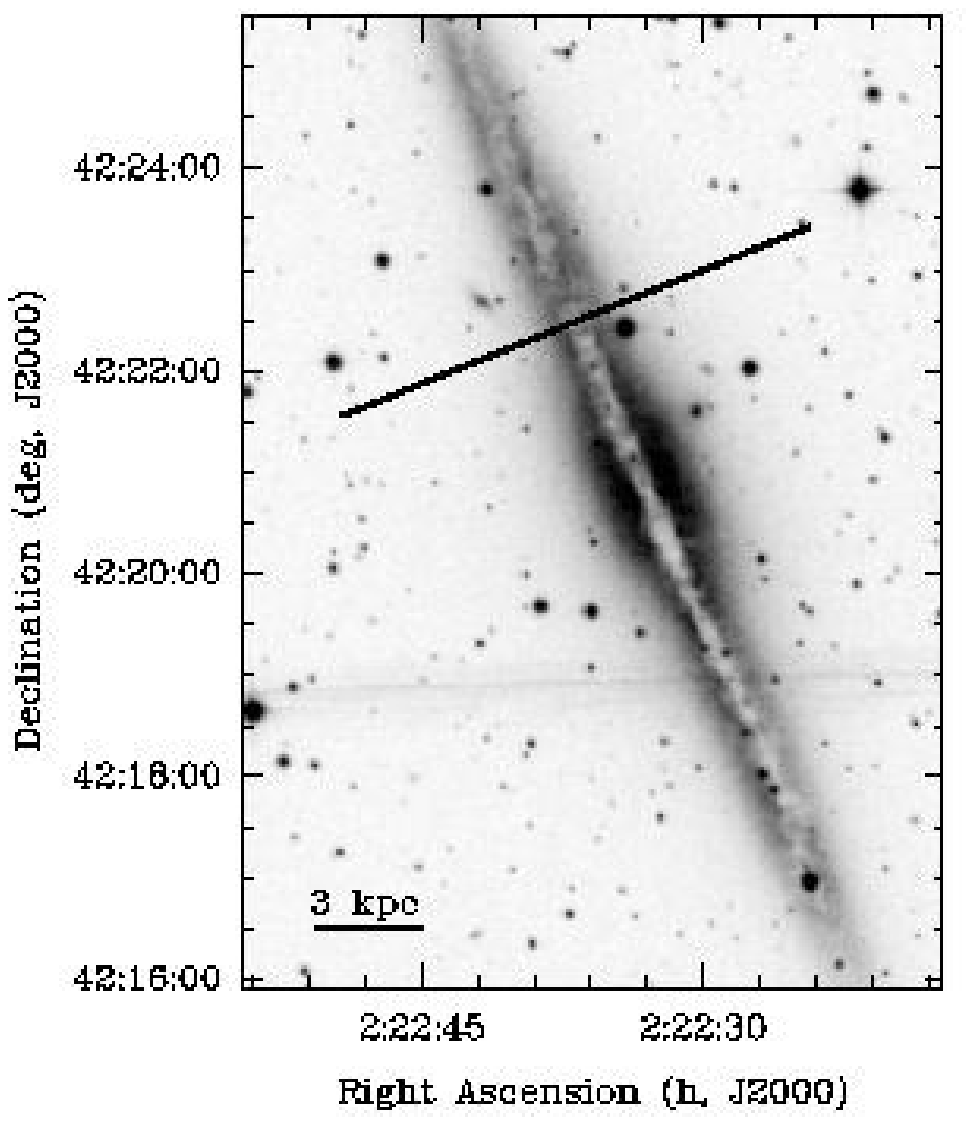}
\caption{NGC\,891 taken from the Digitized Sky Survey (Second Generation).
The position of the slit is shown. In the following plots of NGC\,891, positive
numbers of parsecs refer to the right part of the slit counting westward,
negative numbers refer to the left part of the slit counting eastward. We
assumed a distance of 9.6\,Mpc.}
\end{figure}
\begin{figure}
\plotone{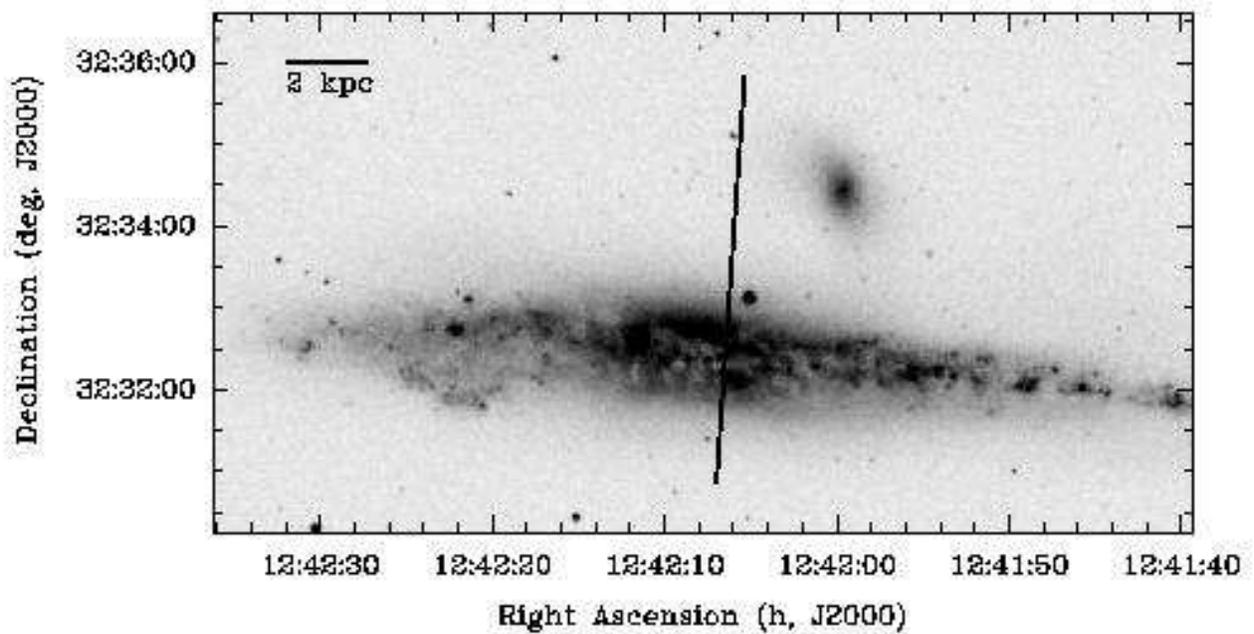}
\caption{NGC\,4631 taken from the Digitized Sky Survey (Second Generation).
The position of the slit is shown. In the following plots of NGC\,4631, positive
numbers of parsecs refer to the upper part of the slit counting northward,
negative numbers refer to the lower part of the slit counting southward. We
assumed a distance of 6.9\,Mpc.}
\end{figure}
\begin{figure}
\epsscale{0.80}
\plotone{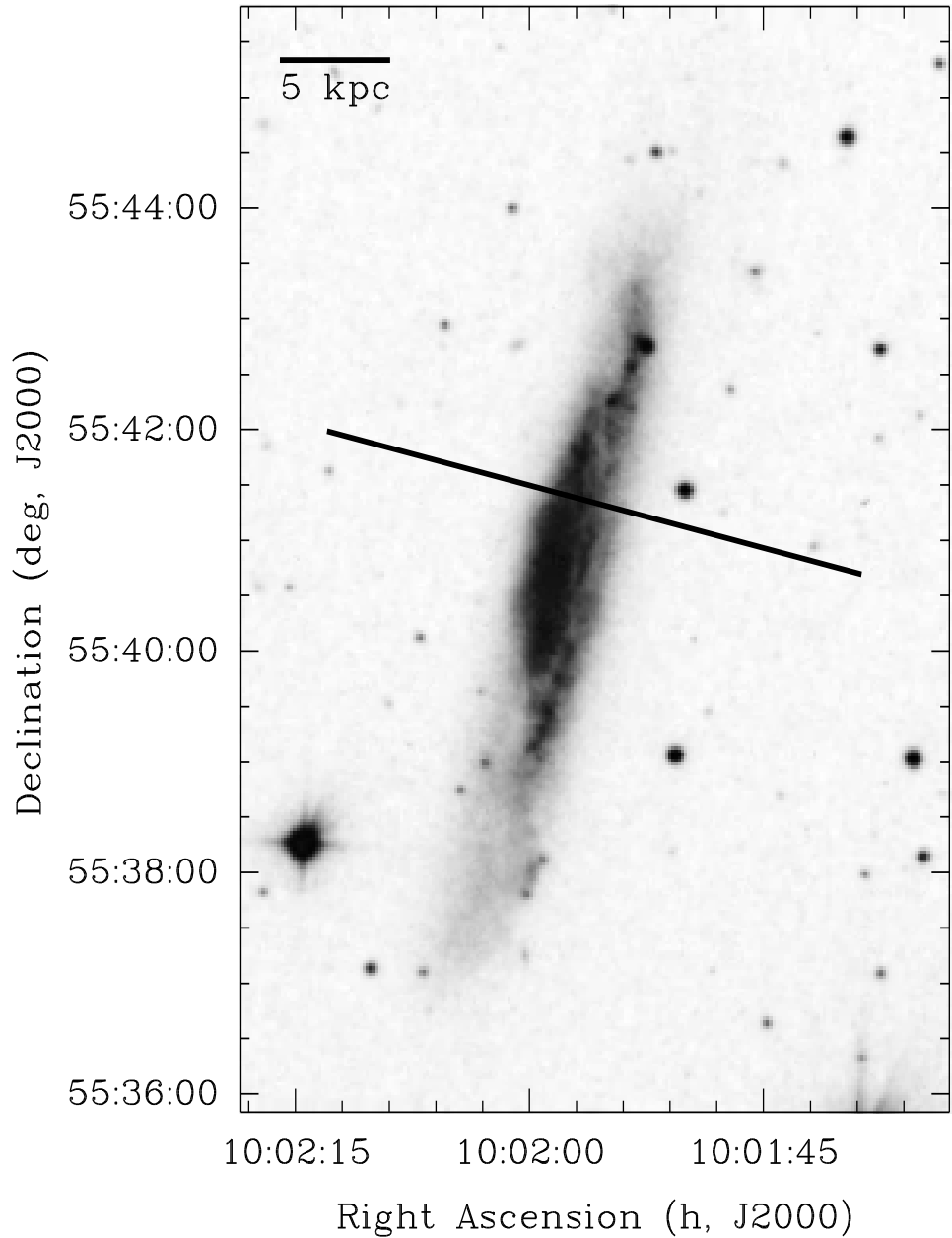}
\caption{NGC\,3079 taken from the Digitized Sky Survey (First Generation).
The position of the slit is shown. In the following plots of NGC\,3079, positive
numbers of parsecs refer to the left part of the slit counting eastward,
negative numbers refer to the right part of the slit counting westward. We
assumed a distance of 17.3\,Mpc.}
\end{figure}
\begin{figure}
\plotone{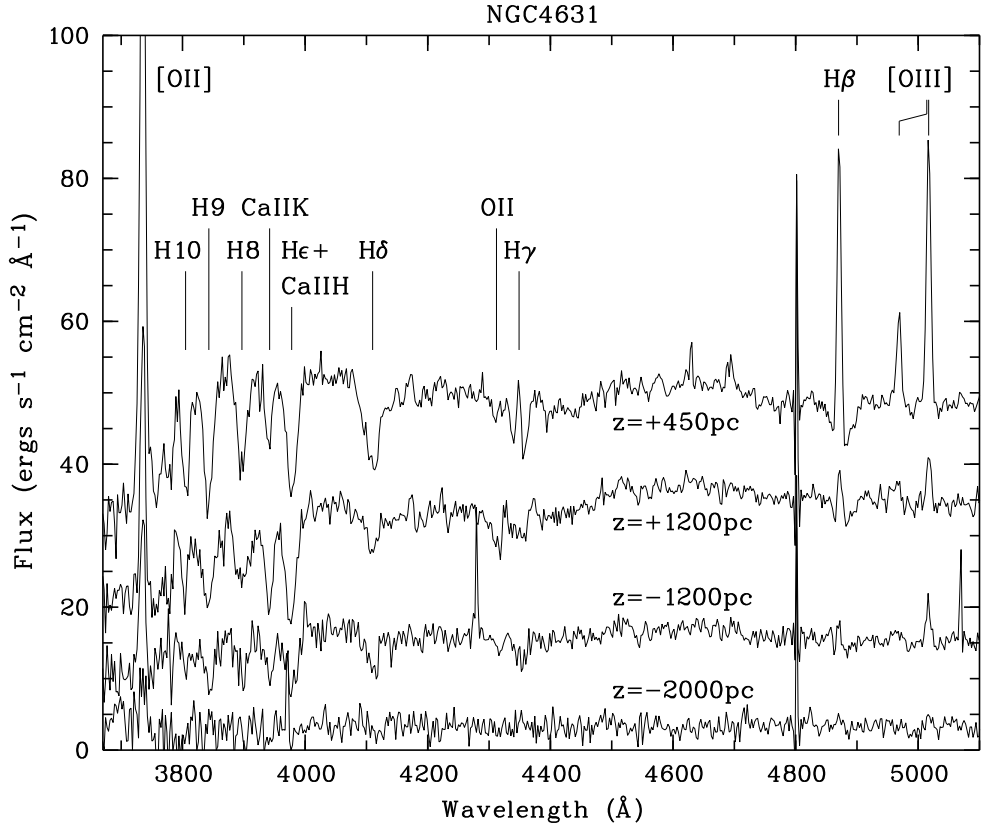}
\caption{Blue part of the spectrum of NGC\,4631. Four different rows of the
longslit spectrum are shown. Each spectrum is an average of nine rows. H$\gamma$
emission can clearly be seen at $z=+450$\,pc inside the H$\gamma$ absorption
line. The H$\gamma$ emission disappears in the noise at $|z|=1200$\,pc. H$\beta$
emission can still be seen at $z=-2000$\,pc. The spectra also show that stellar
continuum is clearly present at $|z|=1200$\,pc.}
\end{figure}
\begin{figure}
\plotone{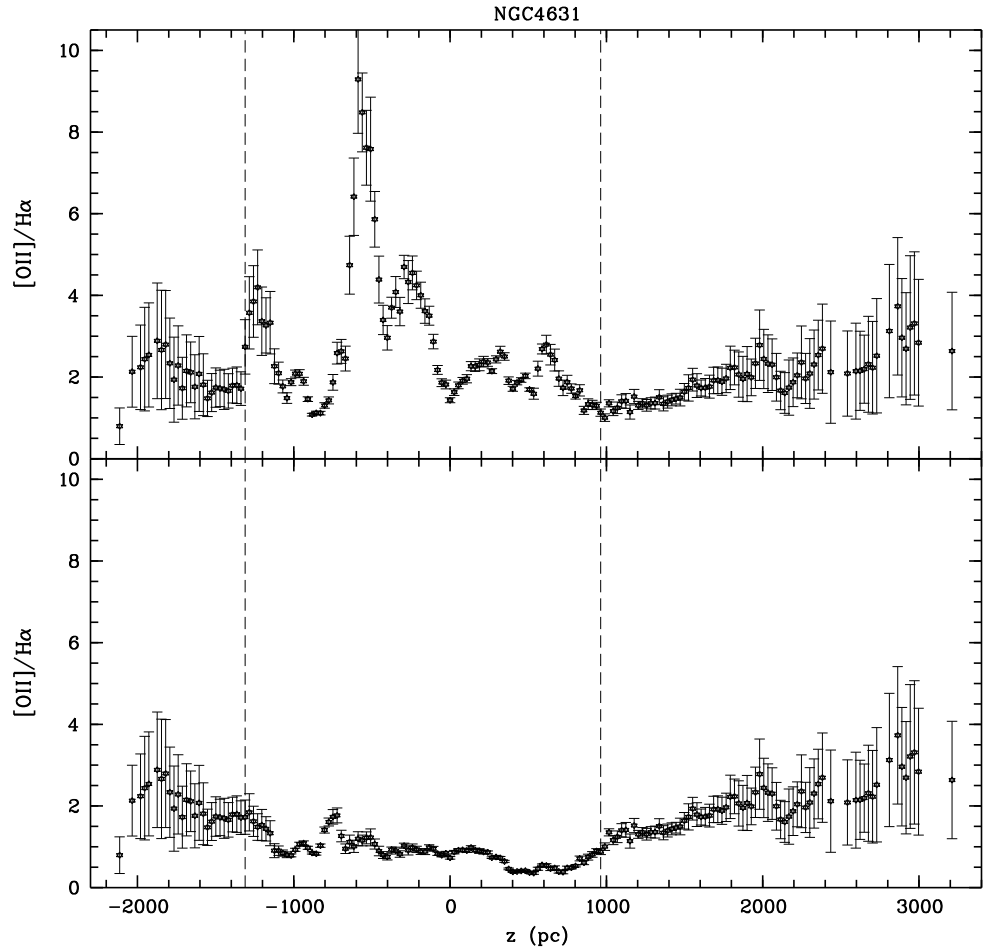}
\caption{[\protect\ion{O}{2}]/H$\alpha$ line ratio in NGC\,4631. {\em Upper panel}:
[\protect\ion{O}{2}]/H$\alpha$ after extinction correction. {\em Lower panel}:
[\protect\ion{O}{2}]/H$\alpha$ without extinction correction. The {\em dashed lines}
show the range affected by the extinction correction. It is not clear how
physical the peaks in the midplane are.}
\end{figure}
\begin{figure}
\epsscale{0.8}
\plotone{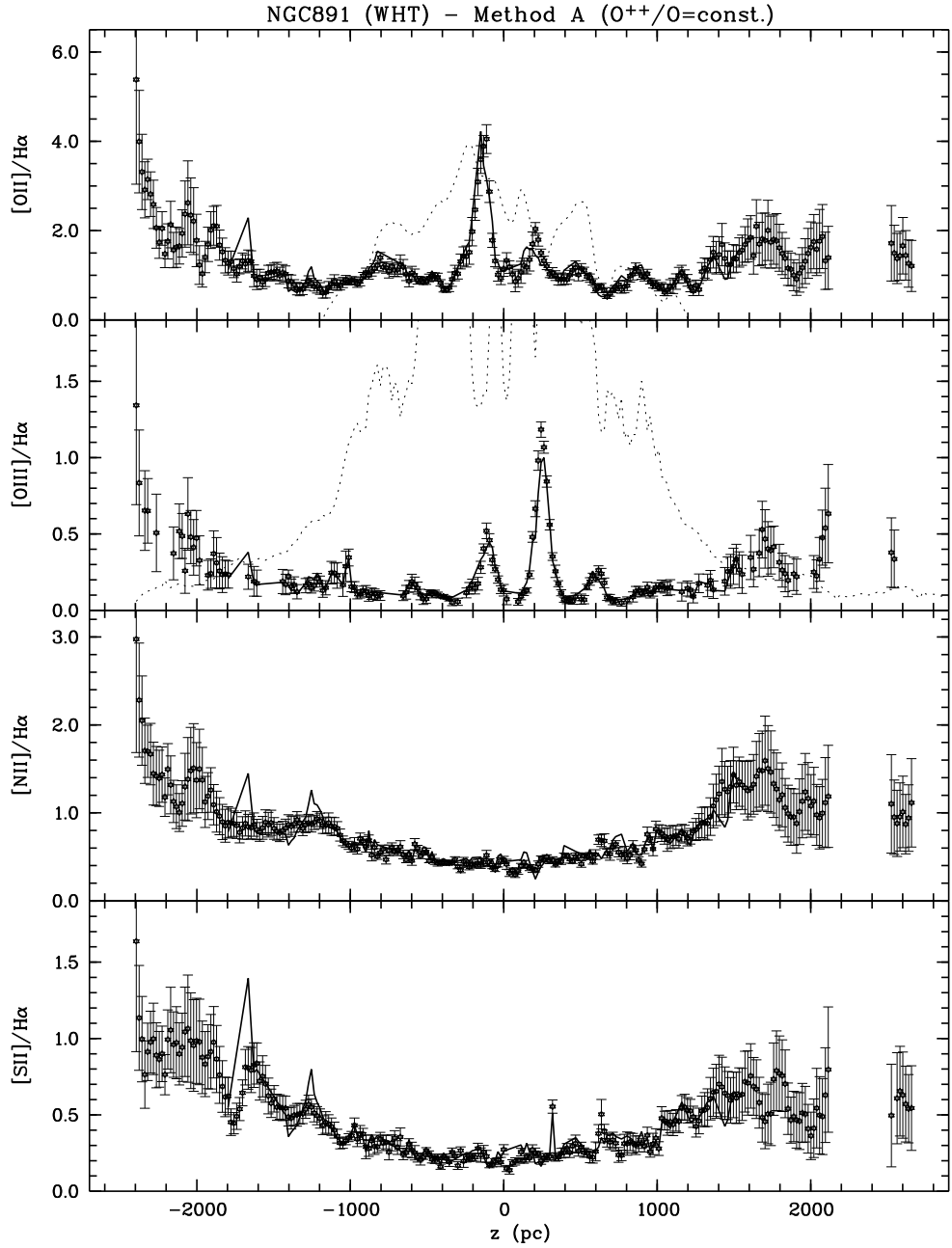}
\caption{H$\alpha$ line ratios and derived properties of NGC\,891 using
Method A (O$^{++}$/O\,=\,const.). ({\em a}) The measured H$\alpha$ line ratios
are compared with the predicted line ratios ({\em solid lines}). Only the
predictions for O$^{++}/{\rm O}=0.15$ are shown. The {\em dotted line} in the
{\em top panel} shows the derived optical depth $\tau$ (unscaled). The {\em
dotted line} in the {\em second from top panel} shows the H$\alpha$ intensity
along the slit scaled down to fit the plot. ({\em b}) The derived electron
temperature, oxygen and nitrogen abundance, and sulfur ionization fraction are
shown for the four different oxygen ionization fractions O$^{++}/{\rm O}=0.05$,
0.10, 0.15, 0.20. The arrows with the attached numbers indicate the maximum
values reached, but not shown in the plot.}
\end{figure}
\begin{figure}
\figurenum{6b}
\plotone{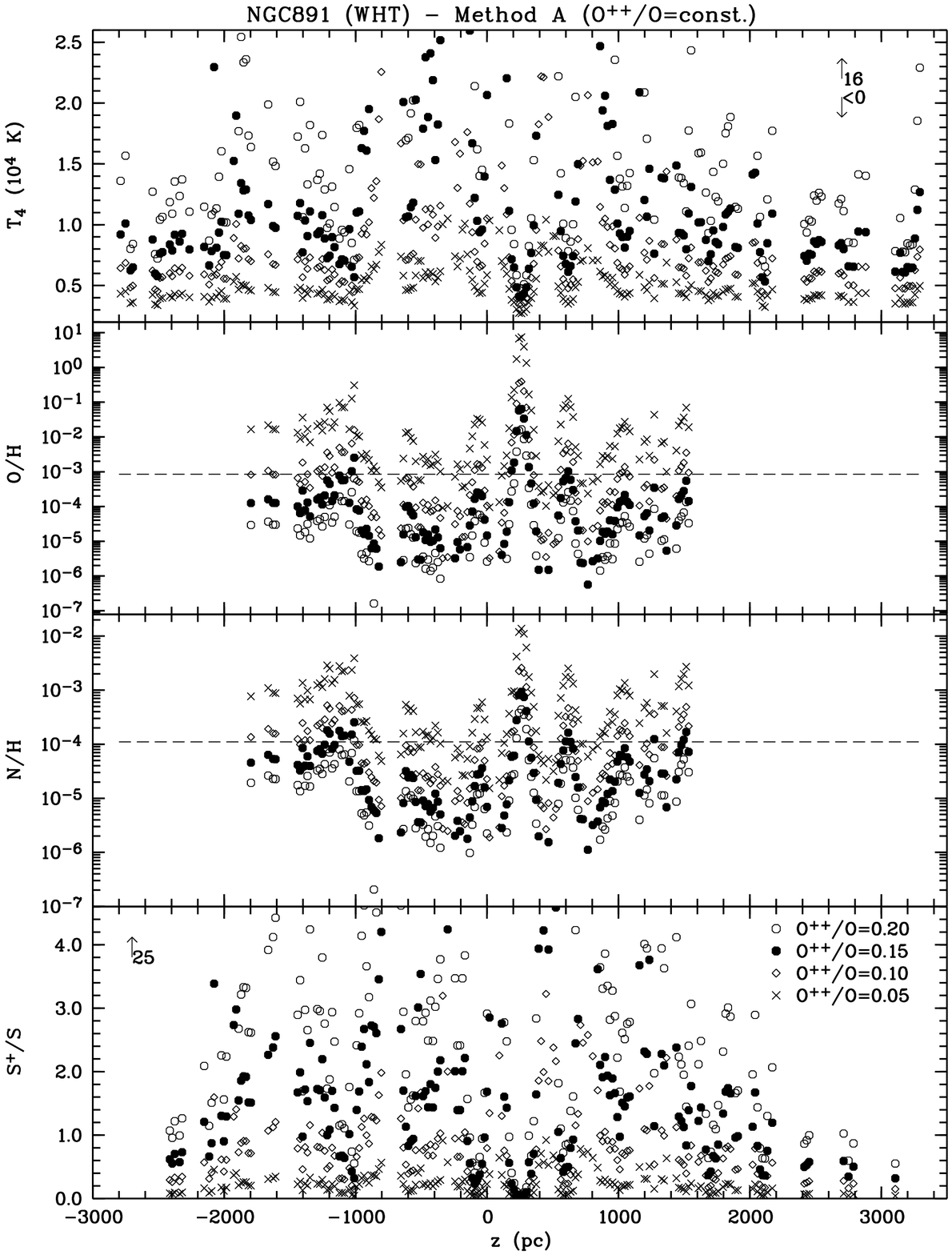}
\caption{}
\end{figure}
\begin{figure}
\plotone{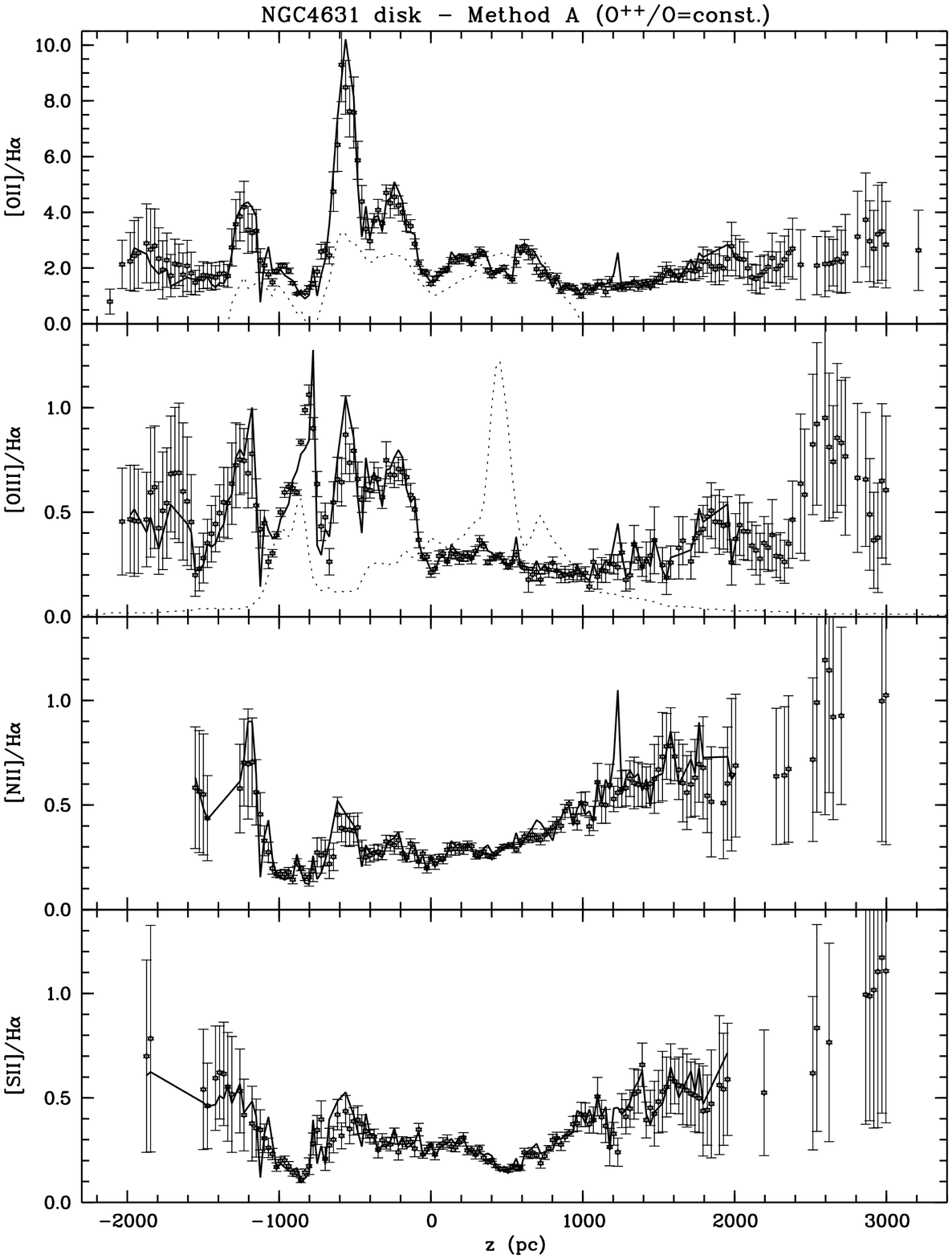}
\caption{Same as Fig. 6, but for NGC\,4631.}
\end{figure}
\begin{figure}
\figurenum{7b}
\plotone{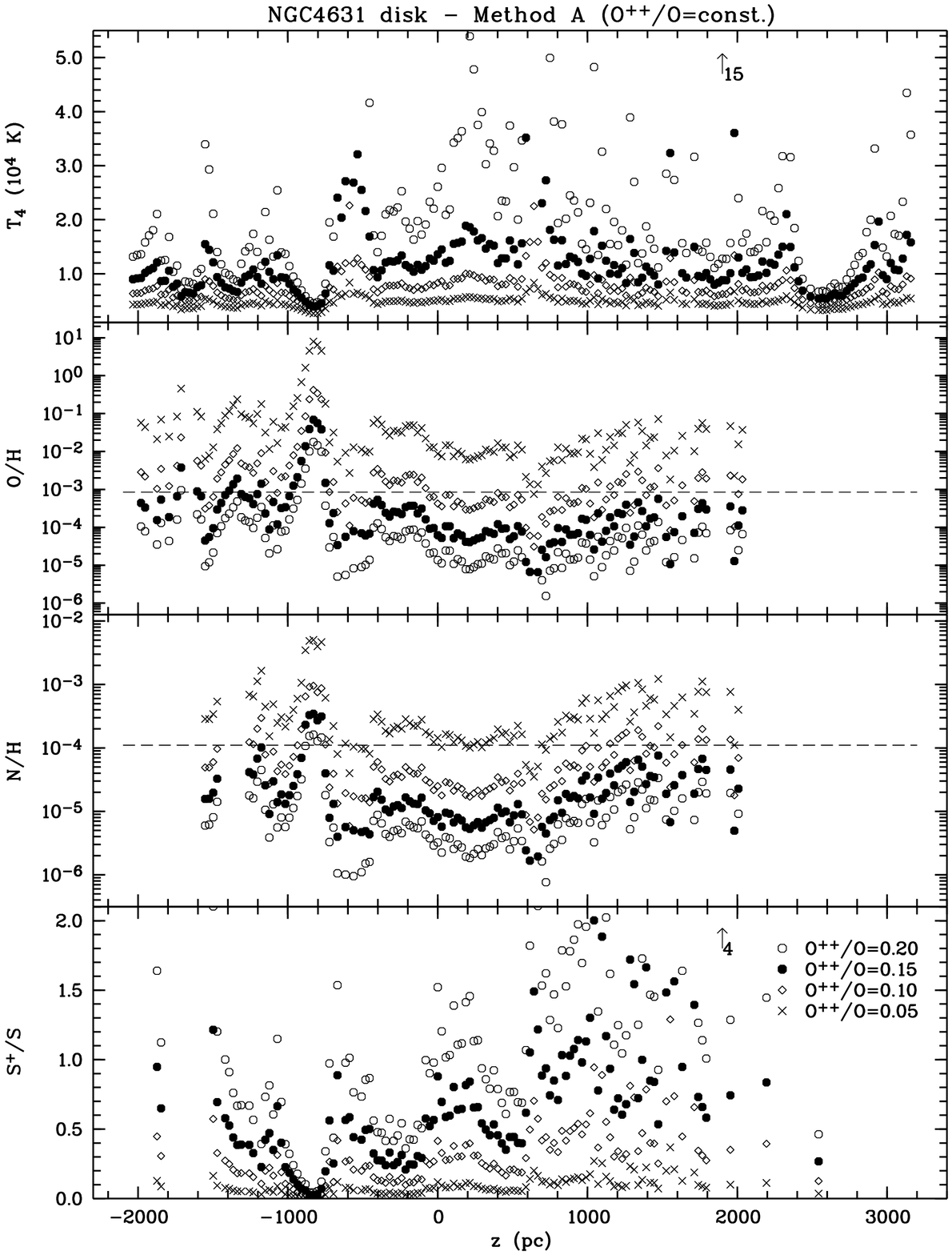}
\caption{}
\end{figure}
\begin{figure}
\plotone{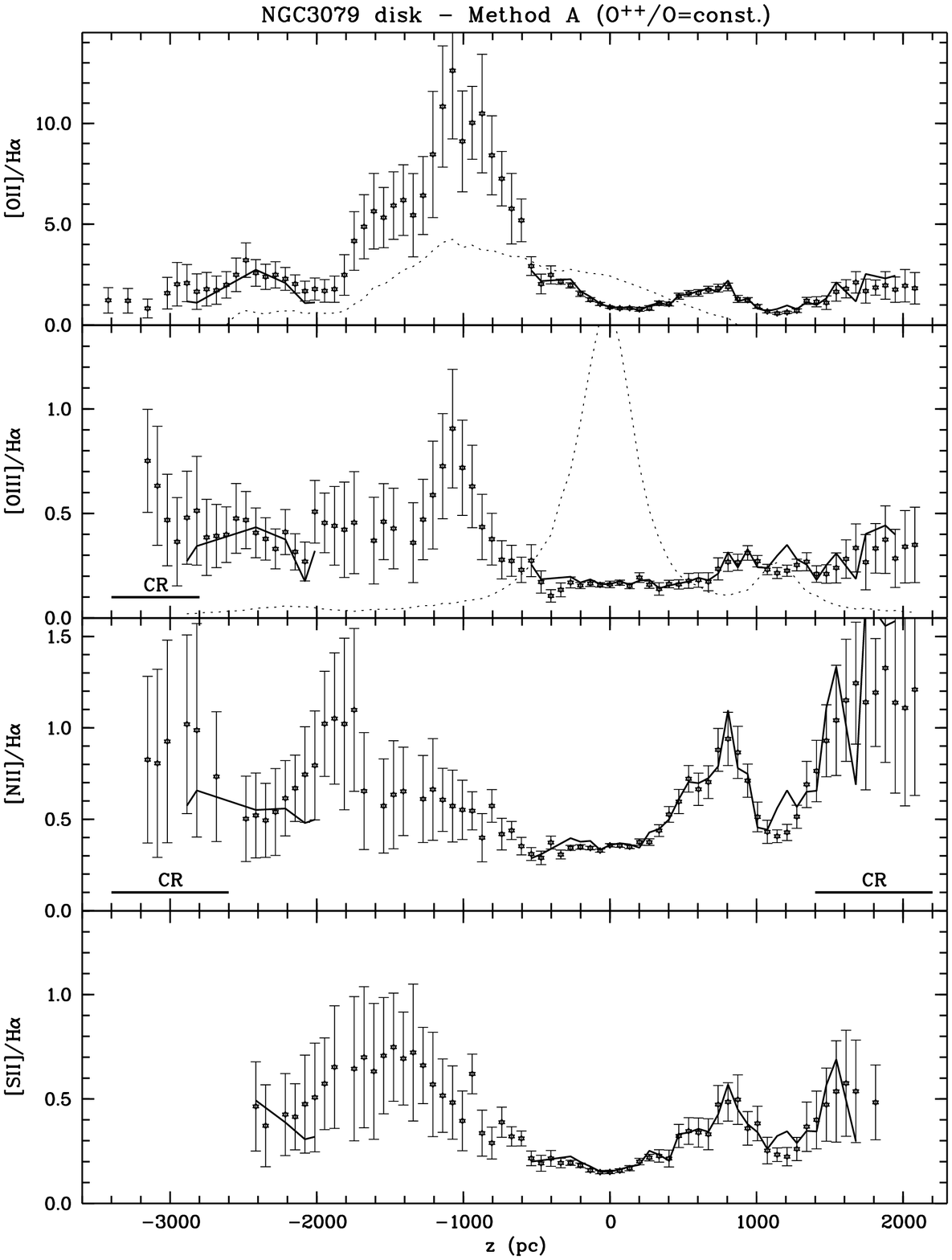}
\caption{Same as Fig. 6, but for NGC\,3079. Data contaminated by cosmic ray
hits are marked with the bars labeled ``CR''.}
\end{figure}
\begin{figure}
\figurenum{8b}
\plotone{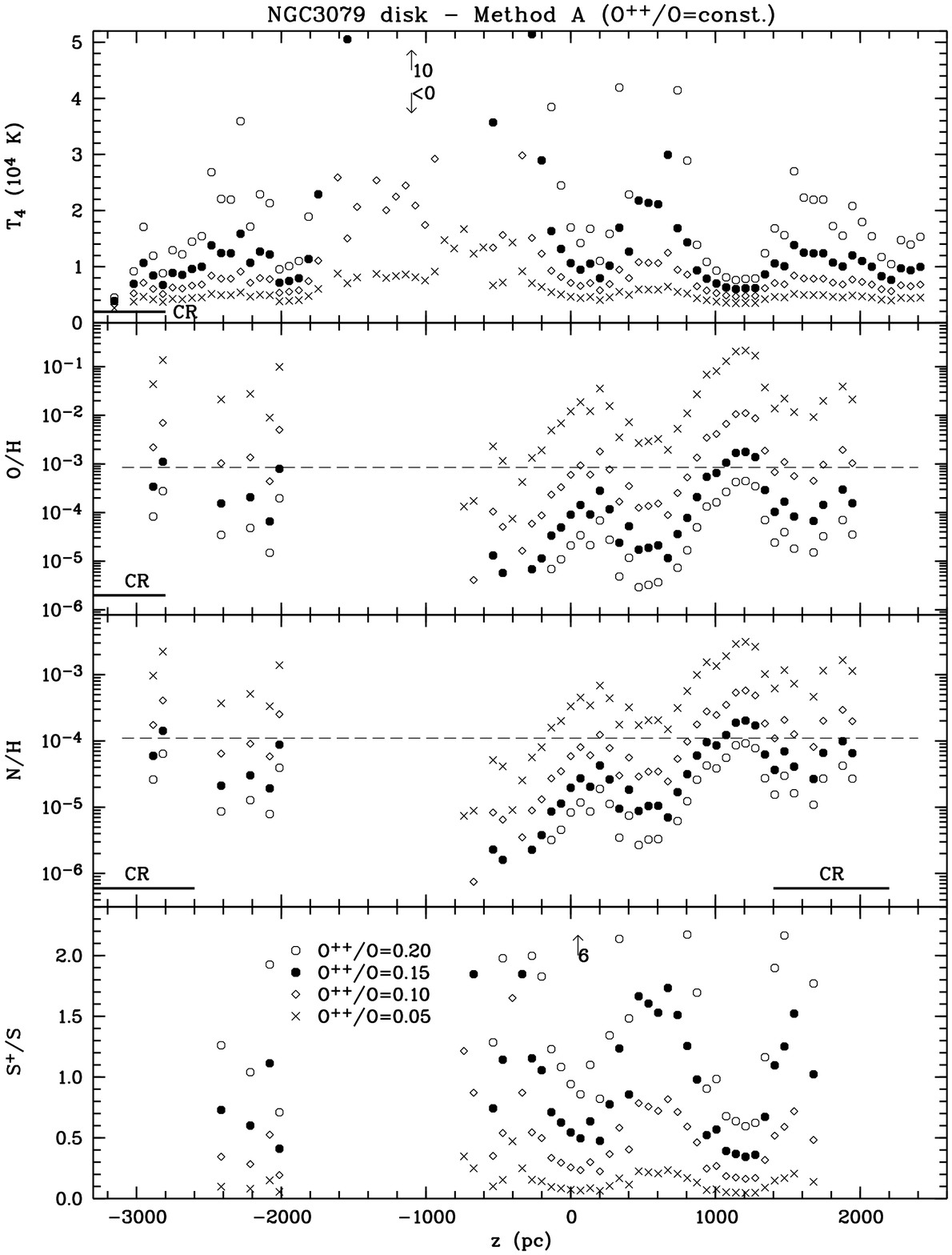}
\caption{}
\end{figure}
\begin{figure}
\plotone{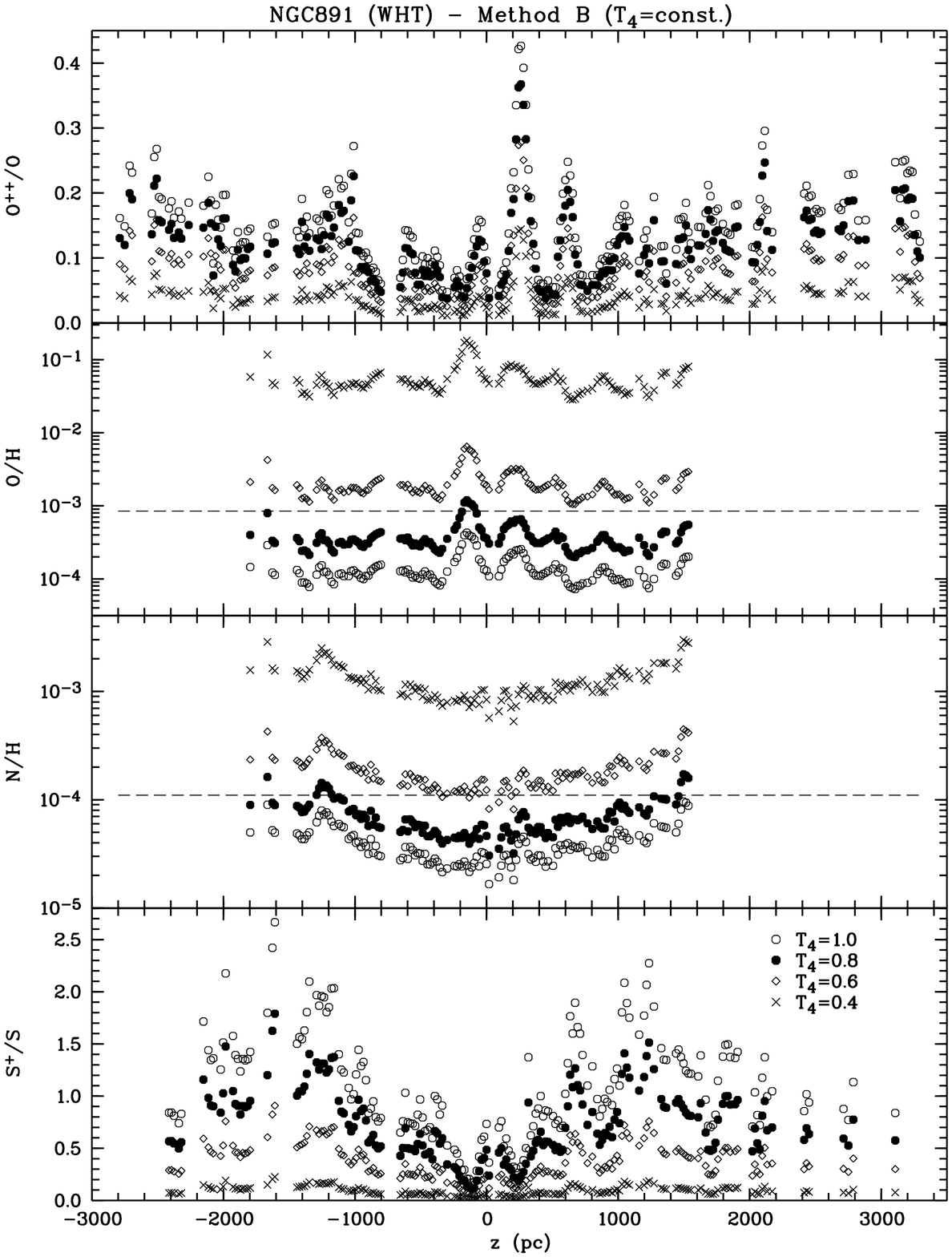}
\caption{Derived properties of NGC\,891 using Method B ($T_4$\,=\,const.).
The derived ionization fraction of doubly ionized oxygen, the oxygen and
nitrogen abundance, and the sulfur ionization fraction are shown for the four
different temperatures $T_4=0.4$, 0.6, 0.8, 1.0 ($T_4$ measured in 10\,000\,K).}
\end{figure}
\begin{figure}
\plotone{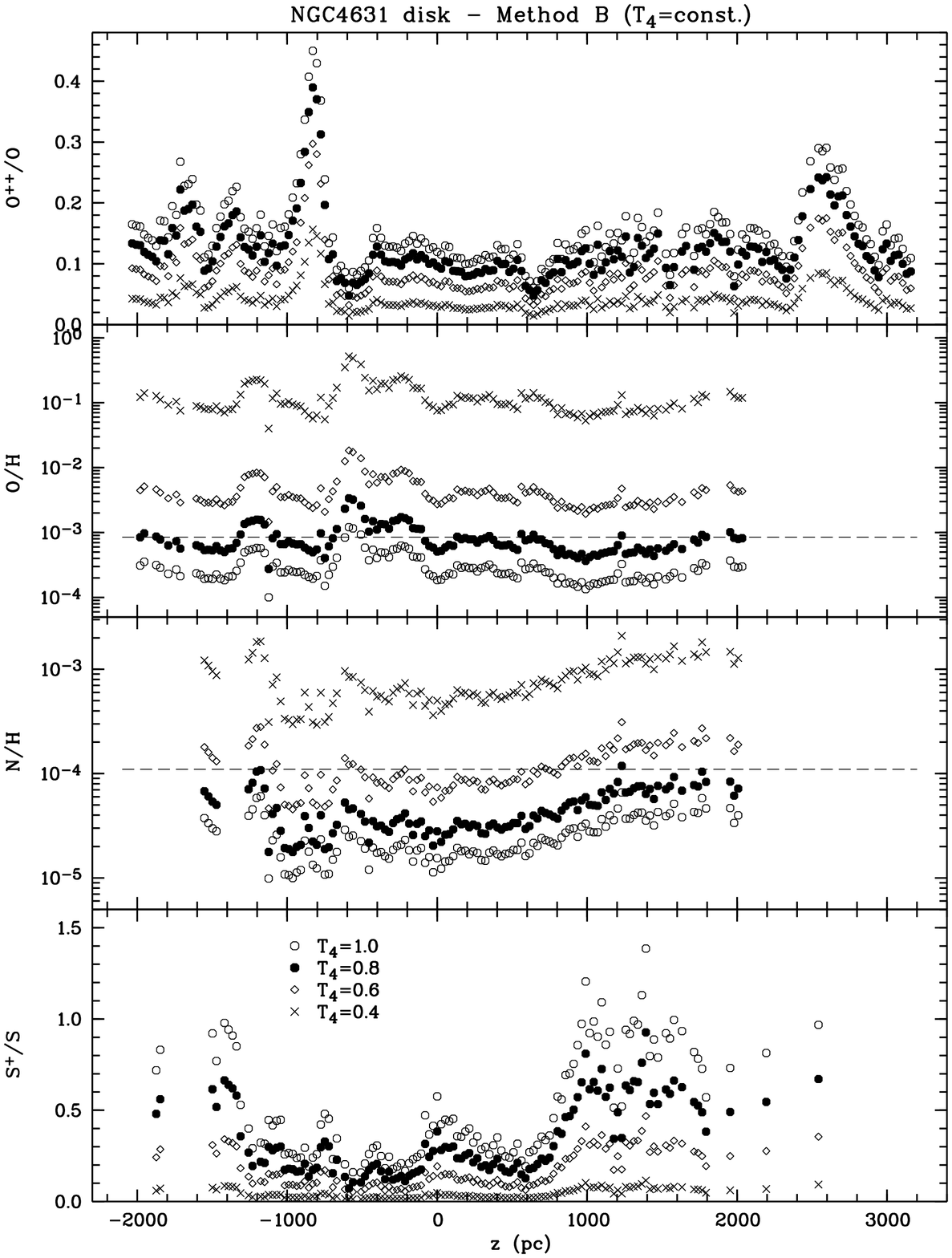}
\caption{Same as Fig. 9, but for NGC\,4631.}
\end{figure}
\begin{figure}
\plotone{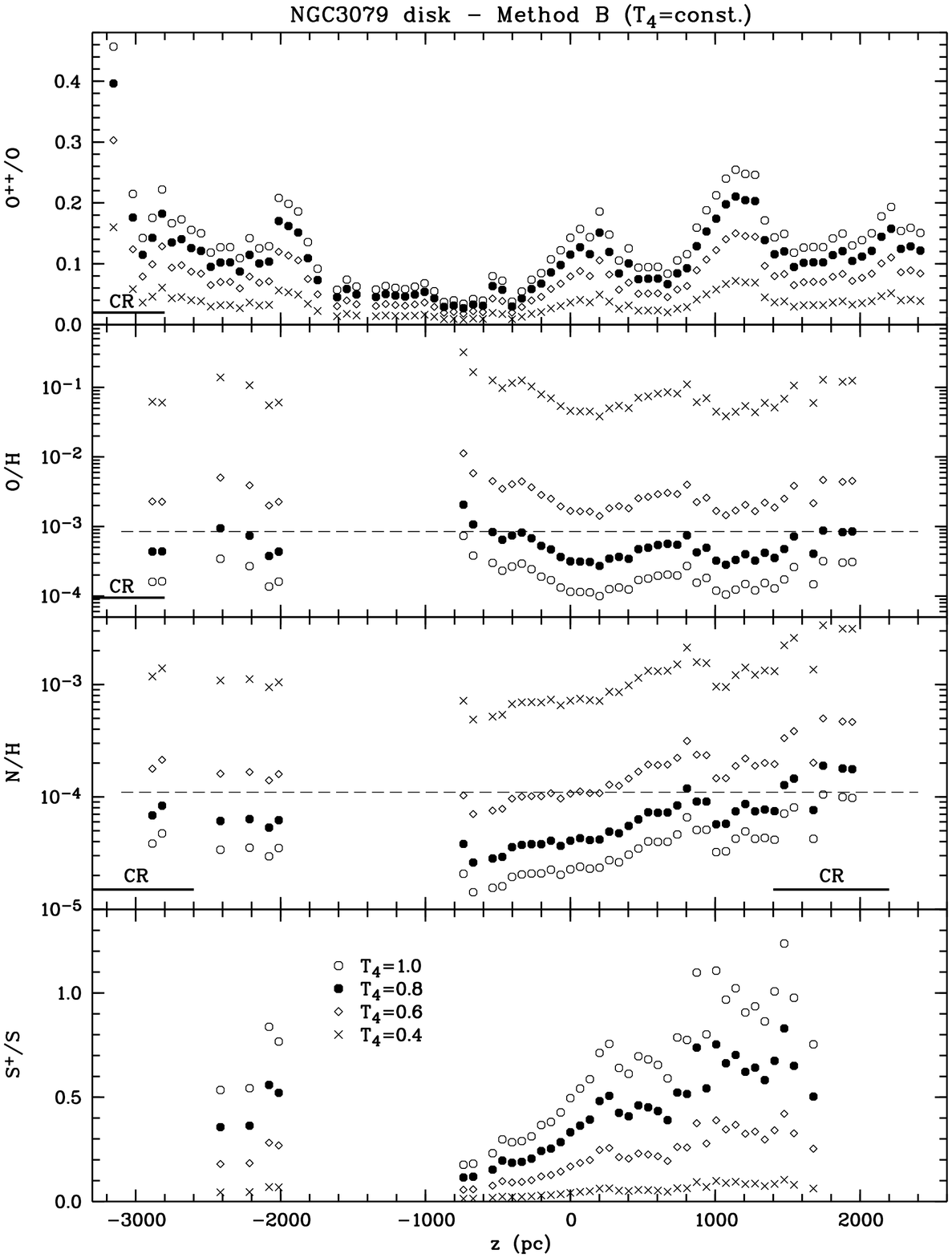}
\caption{Same as Fig. 9, but for NGC\,3079. Data contaminated by cosmic ray
hits are marked with the bars labeled ``CR''.}
\end{figure}
\clearpage
\begin{figure}
\plotone{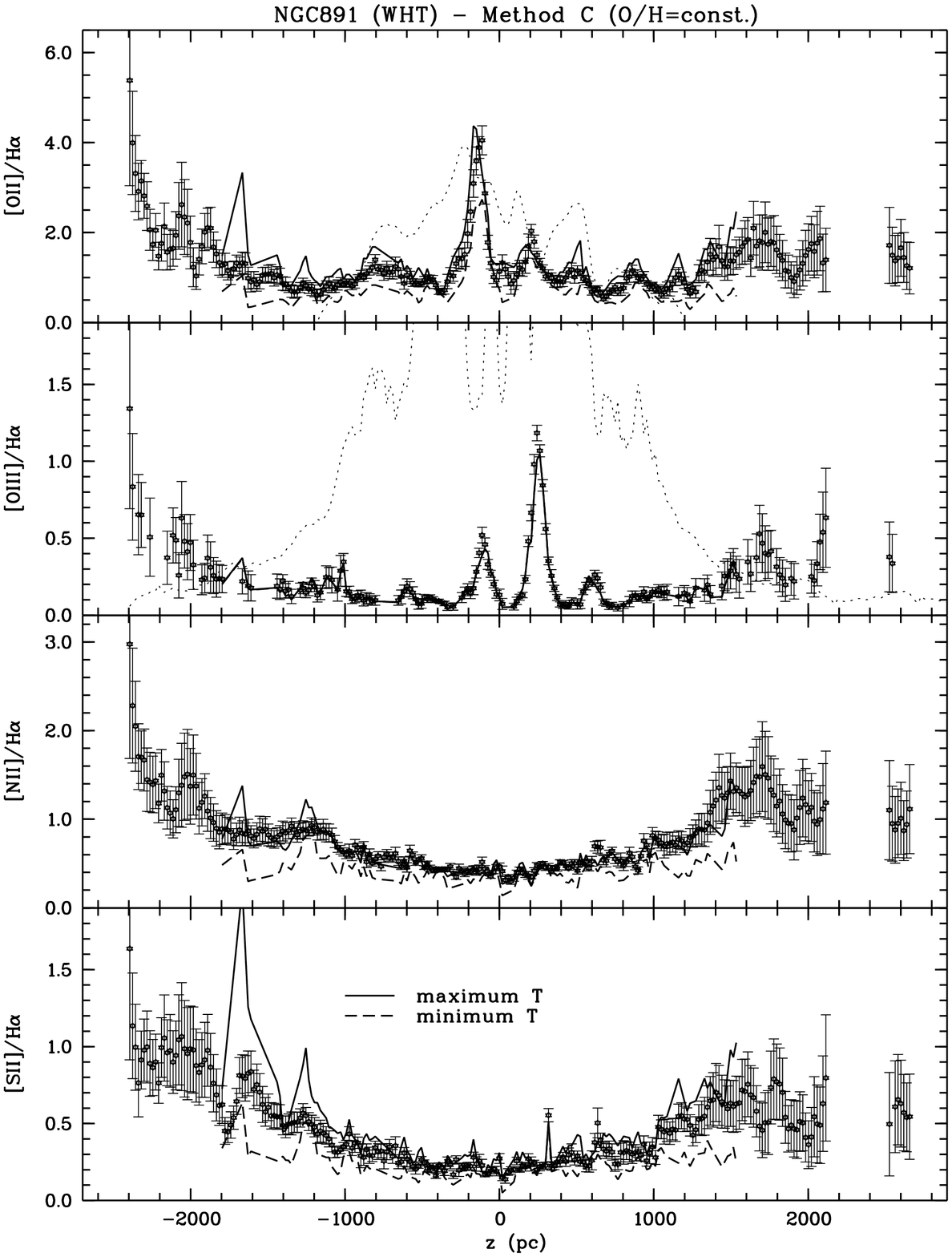}
\caption{H$\alpha$ line ratios and derived properties of NGC\,891 using
Method C (O/H\,=\,const.). ({\em a}) The measured H$\alpha$ line ratios are
compared with the predicted line ratios using the maximum temperature ({\em
solid lines}) and the minimum temperature ({\em dashed lines}). The {\em dotted
line} in the {\em top panel} shows the derived optical depth $\tau$ (unscaled).
The {\em dotted line} in the {\em second from top panel} shows the H$\alpha$
intensity along the slit scaled down to fit the plot. ({\em b}) The derived
electron temperature, ionization fraction of doubly ionized oxygen, the nitrogen
abundance, and the sulfur ionization fraction are shown for both the maximum
temperature ({\em filled circles}) and the minimum temperature ({\em open
diamonds}). ({\em c}) same as ({\em a}), but without extinction correction.
({\em d}) same as ({\em b}), but without extinction correction.}
\end{figure}
\begin{figure}
\figurenum{12b}
\plotone{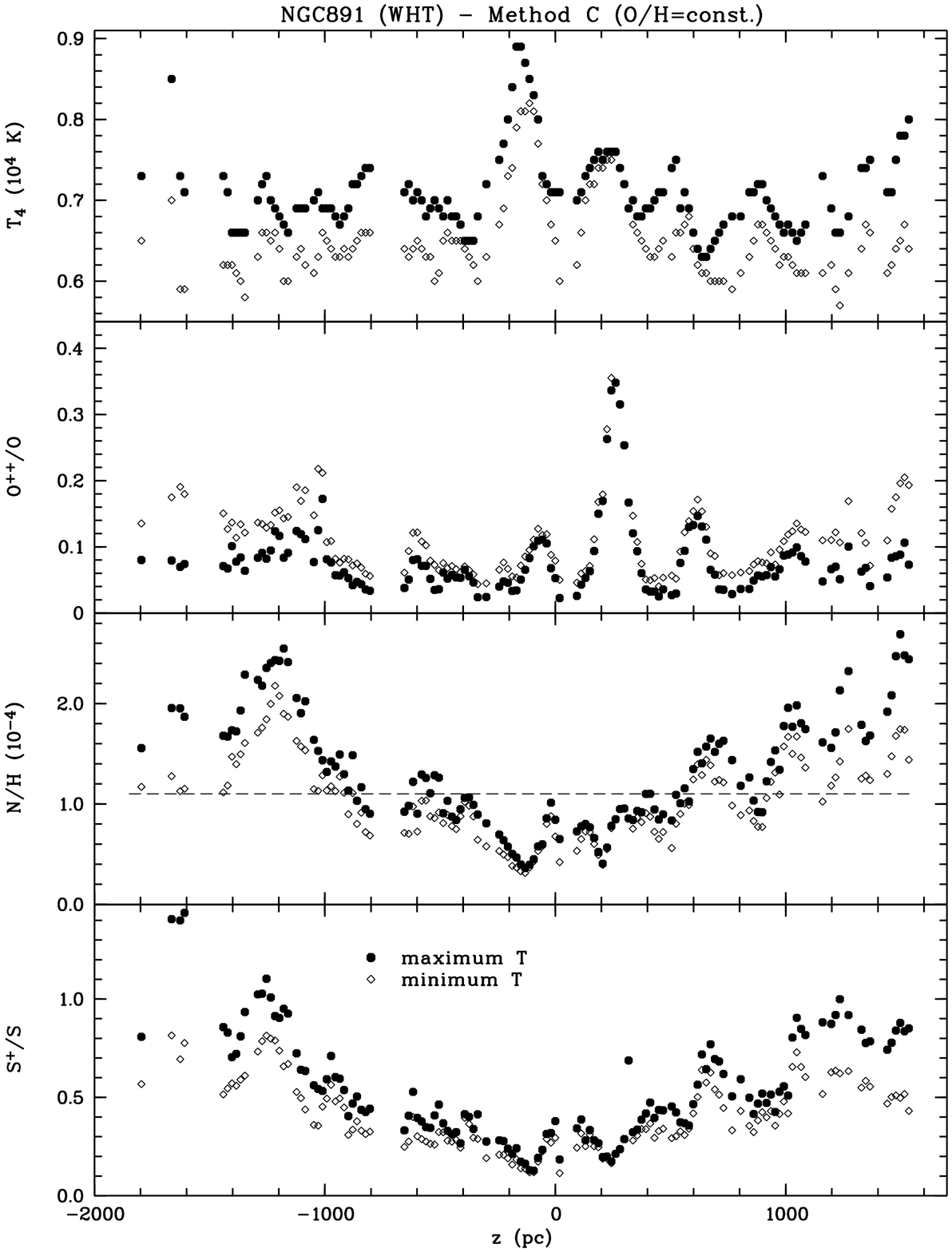}
\caption{}
\end{figure}
\begin{figure}
\figurenum{12c}
\plotone{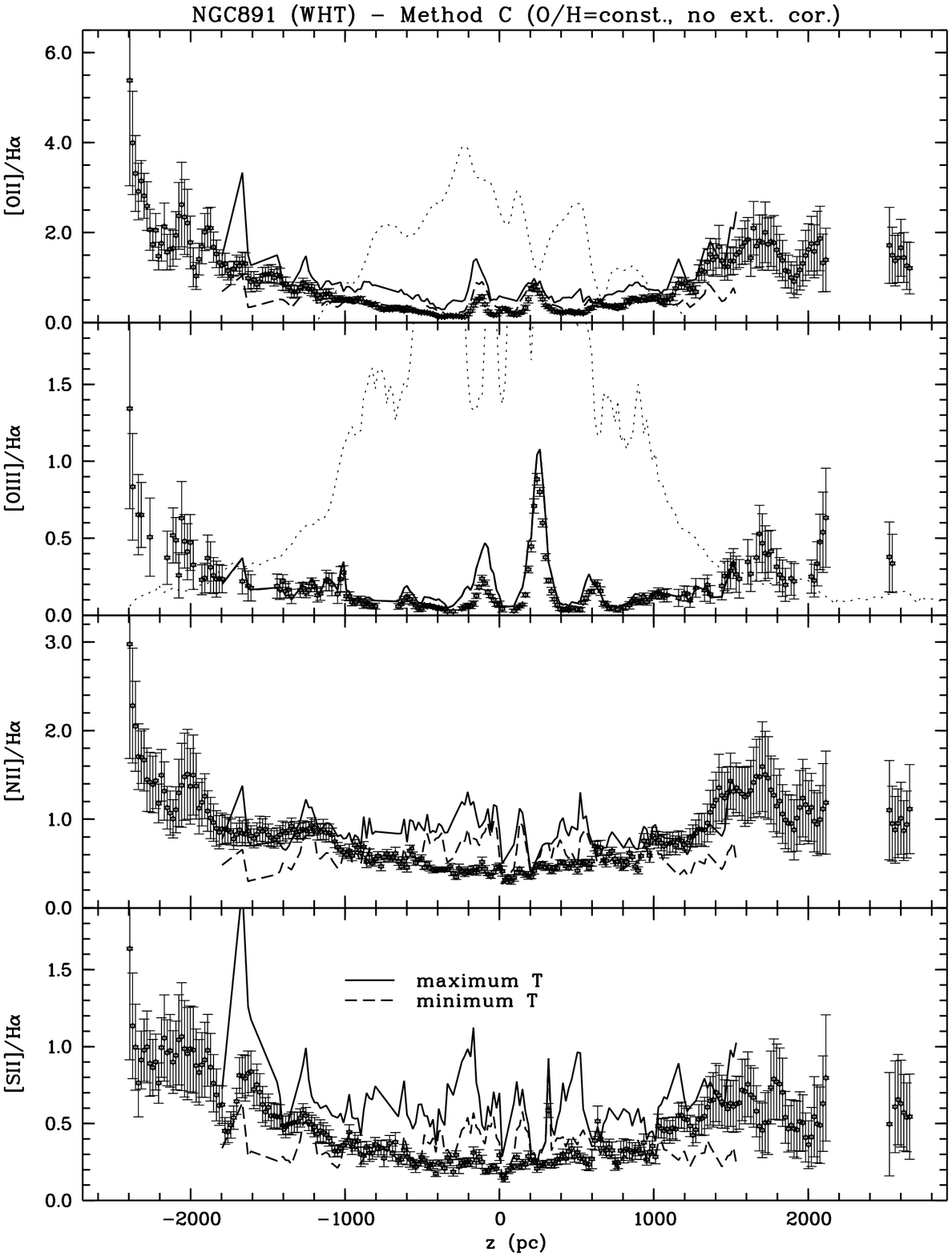}
\caption{}
\end{figure}
\begin{figure}
\figurenum{12d}
\plotone{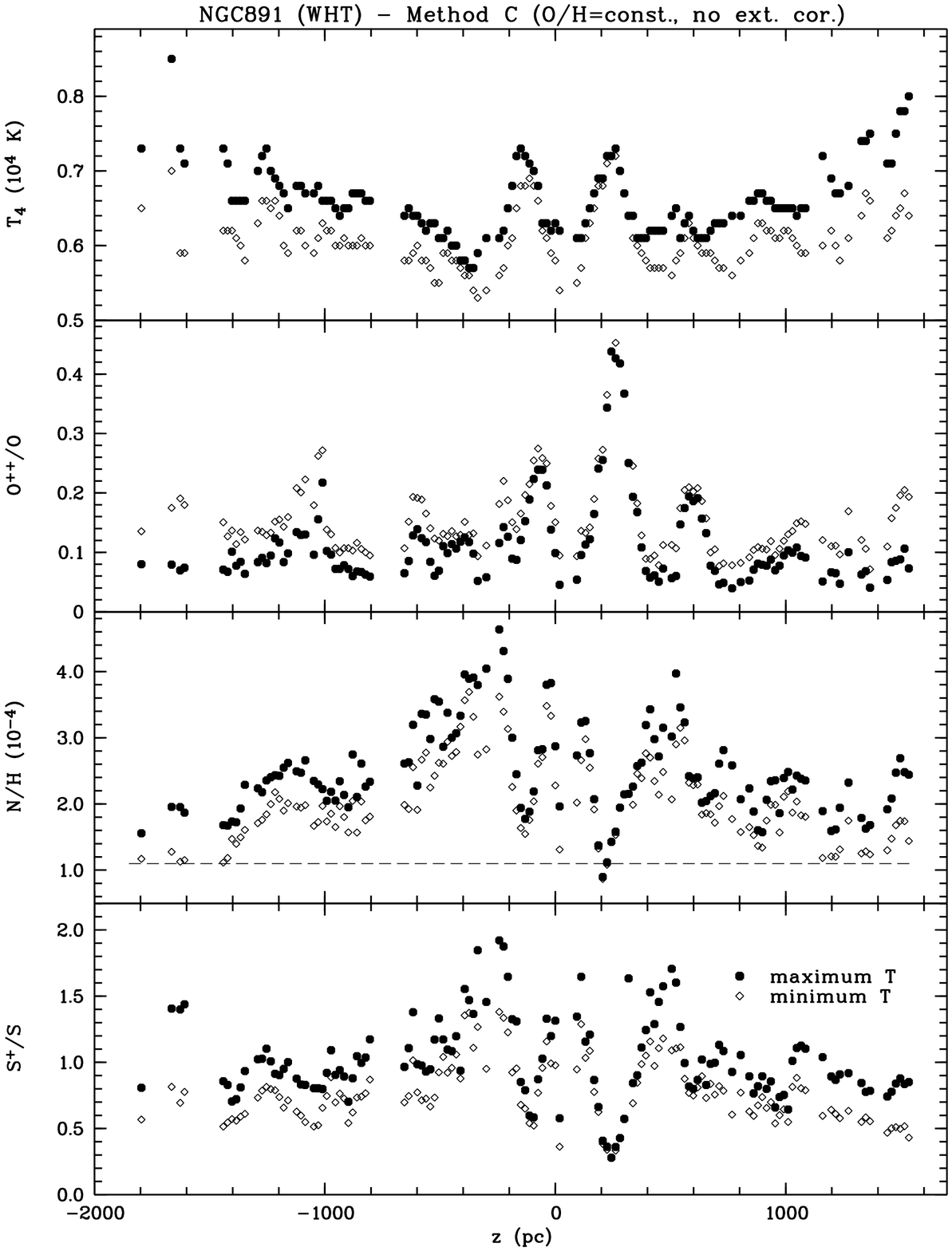}
\caption{}
\end{figure}
\begin{figure}
\plotone{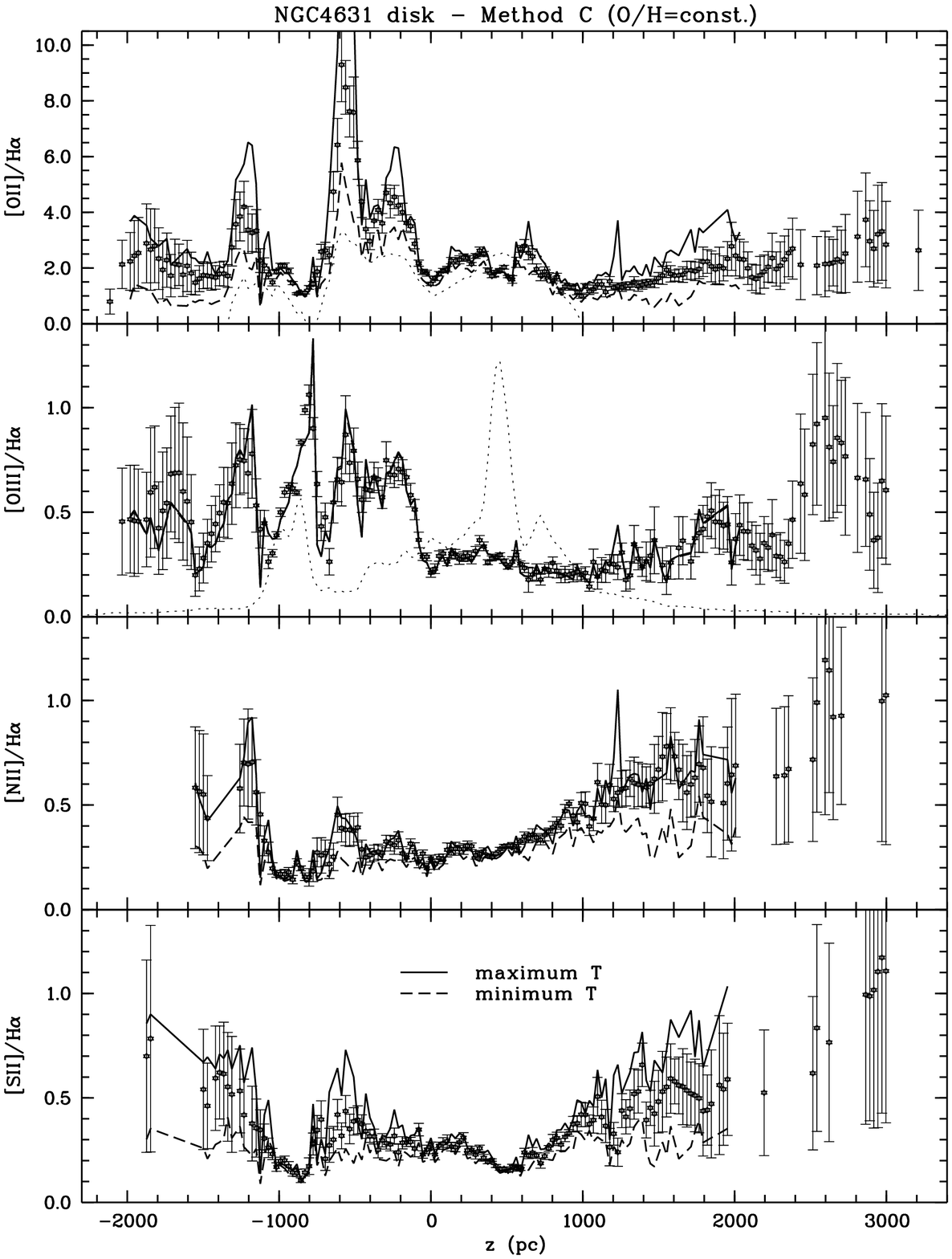}
\caption{Same as Fig. 12, but for NGC\,4631.}
\end{figure}
\begin{figure}
\figurenum{13b}
\plotone{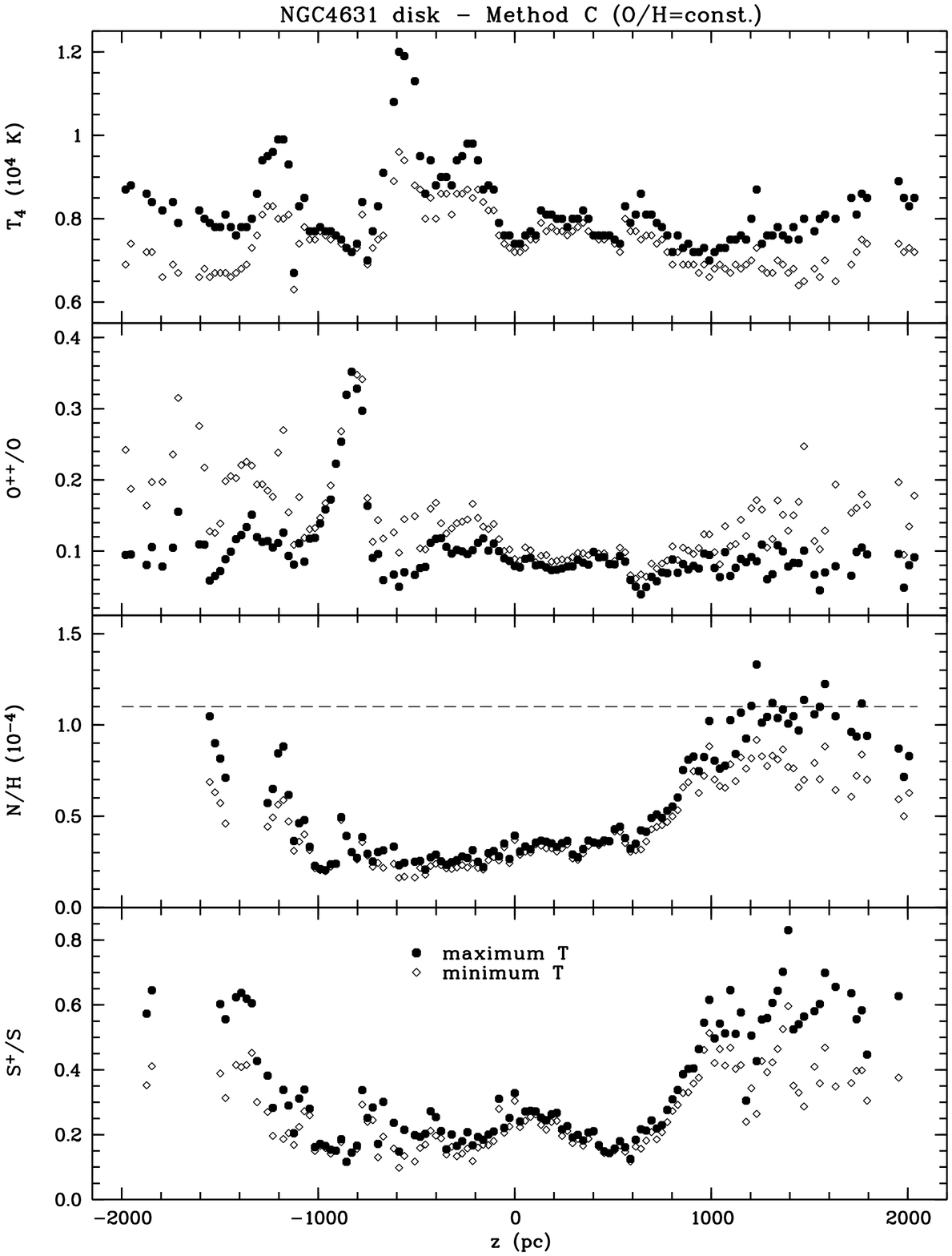}
\caption{}
\end{figure}
\begin{figure}
\figurenum{13c}
\plotone{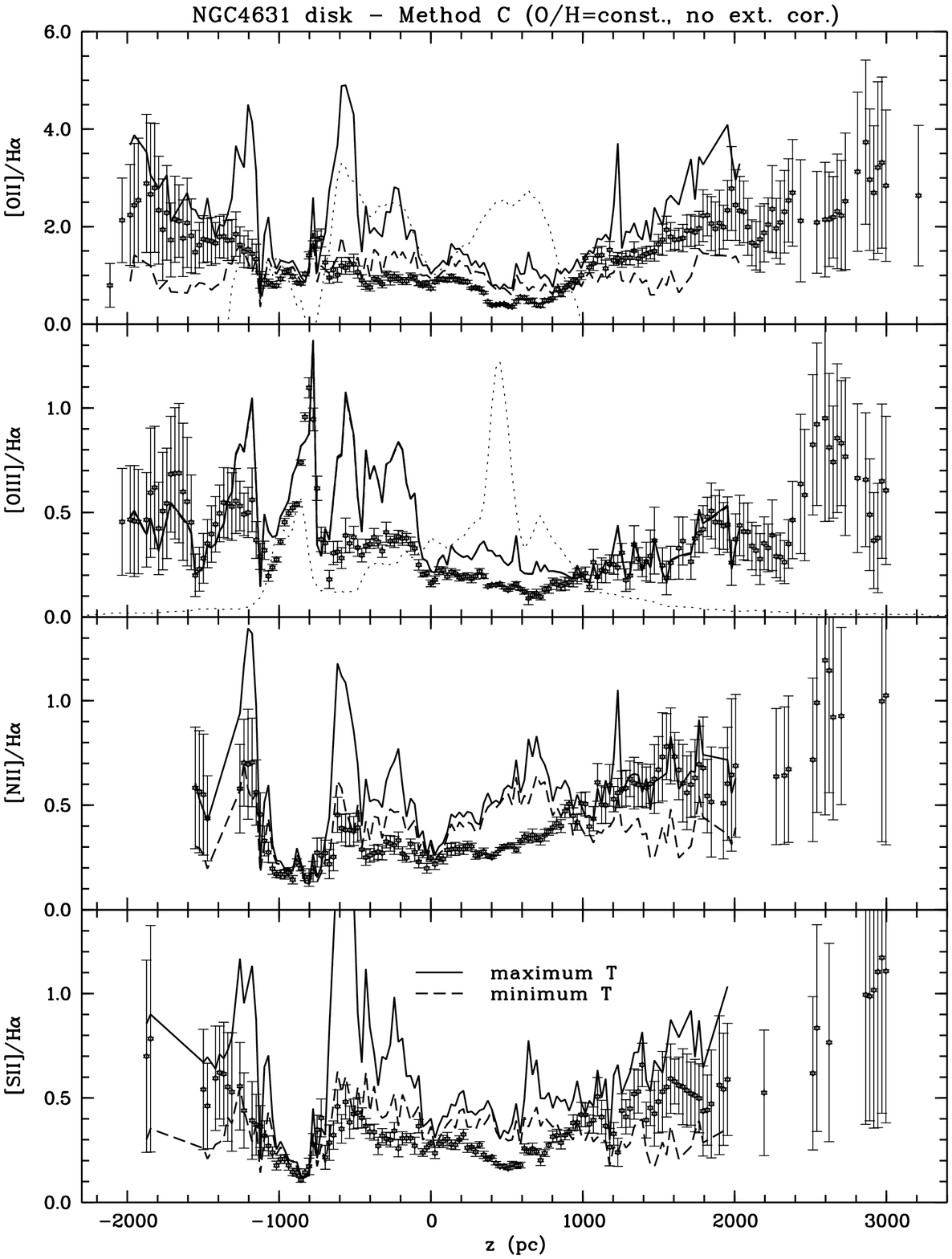}
\caption{}
\end{figure}
\begin{figure}
\figurenum{13d}
\plotone{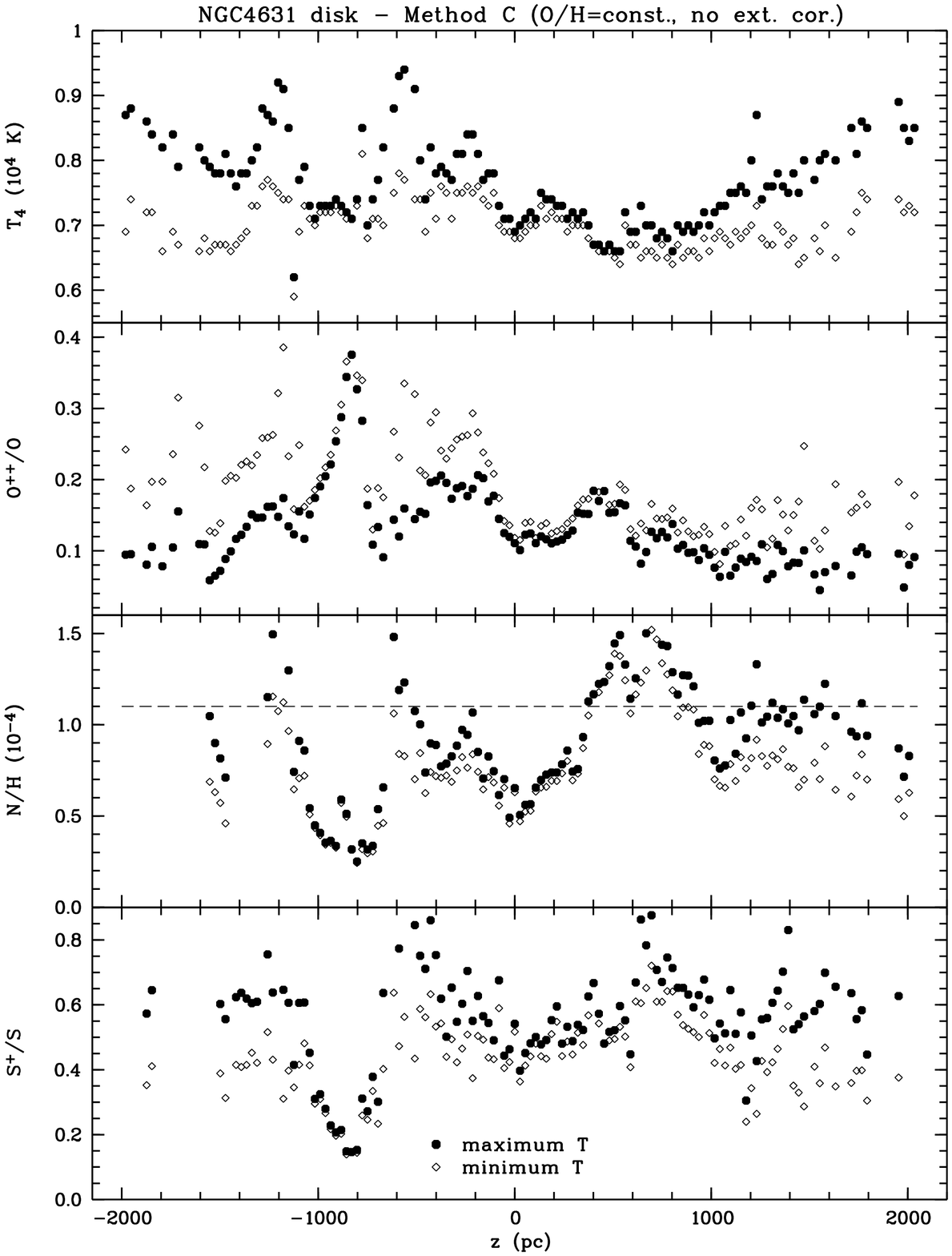}
\caption{}
\end{figure}
\begin{figure}
\plotone{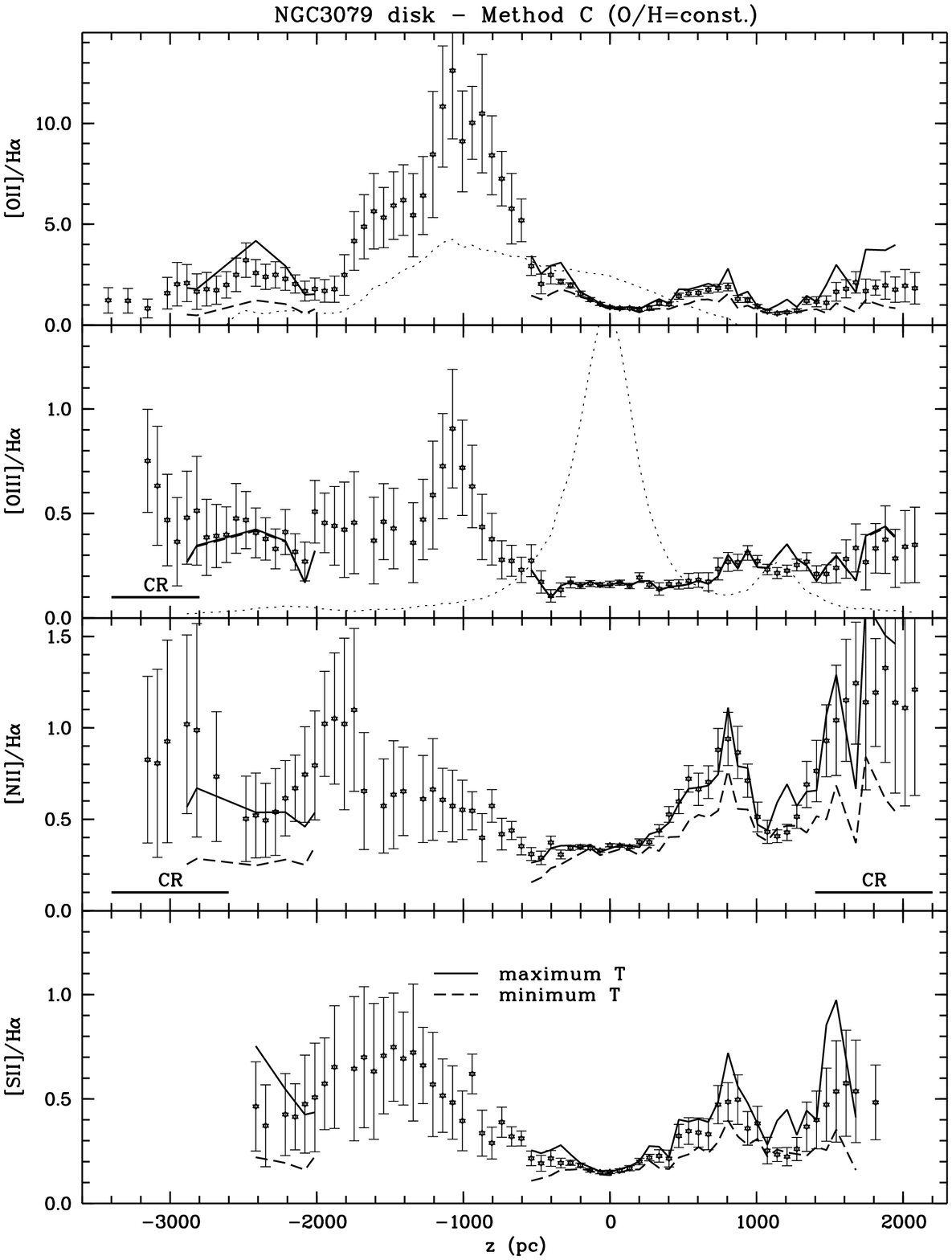}
\caption{Same as Fig. 12, but for NGC\,3079. Data contaminated by cosmic ray
hits are marked with the bars labeled ``CR''.}
\end{figure}
\begin{figure}
\figurenum{14b}
\plotone{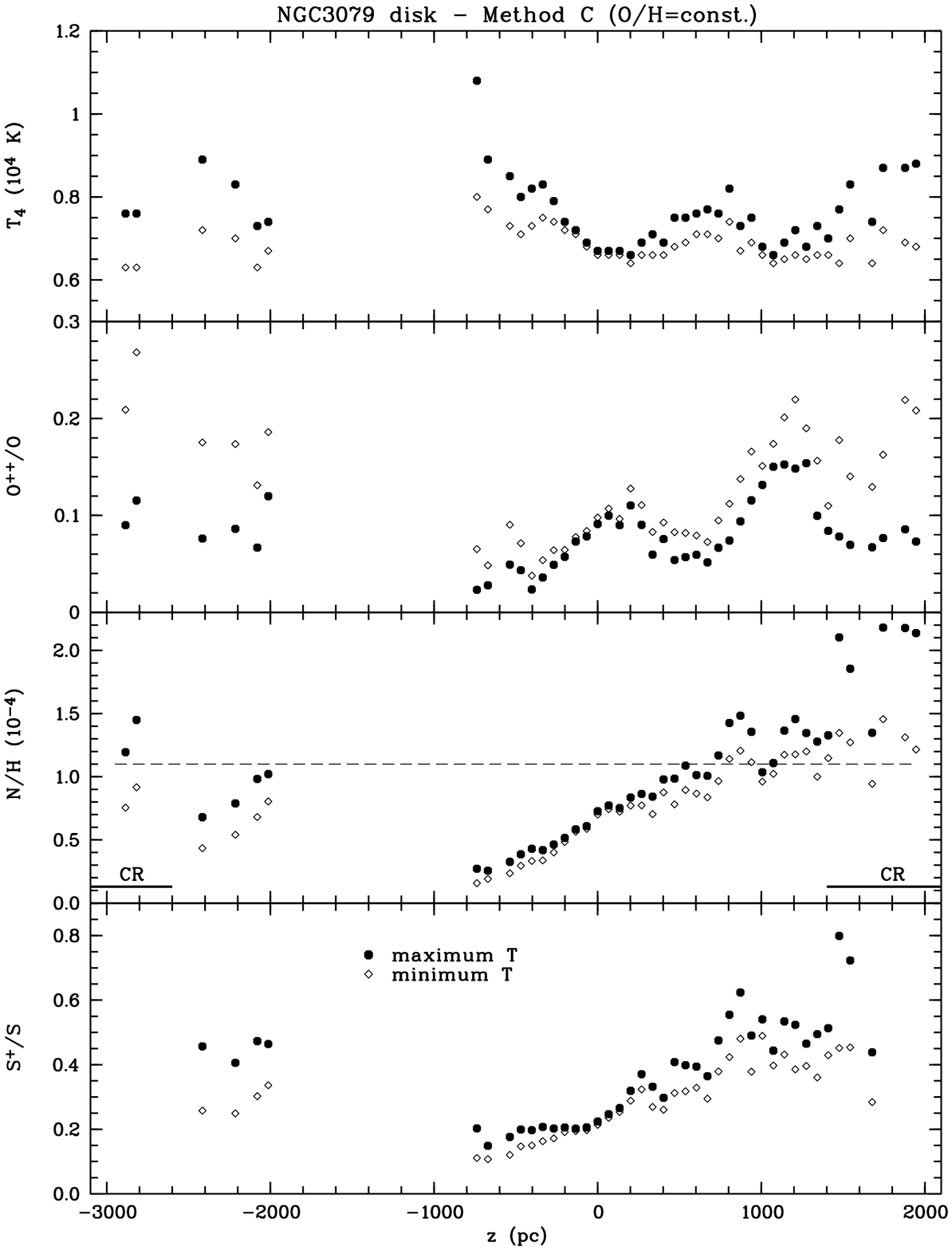}
\caption{}
\end{figure}
\begin{figure}
\figurenum{14c}
\plotone{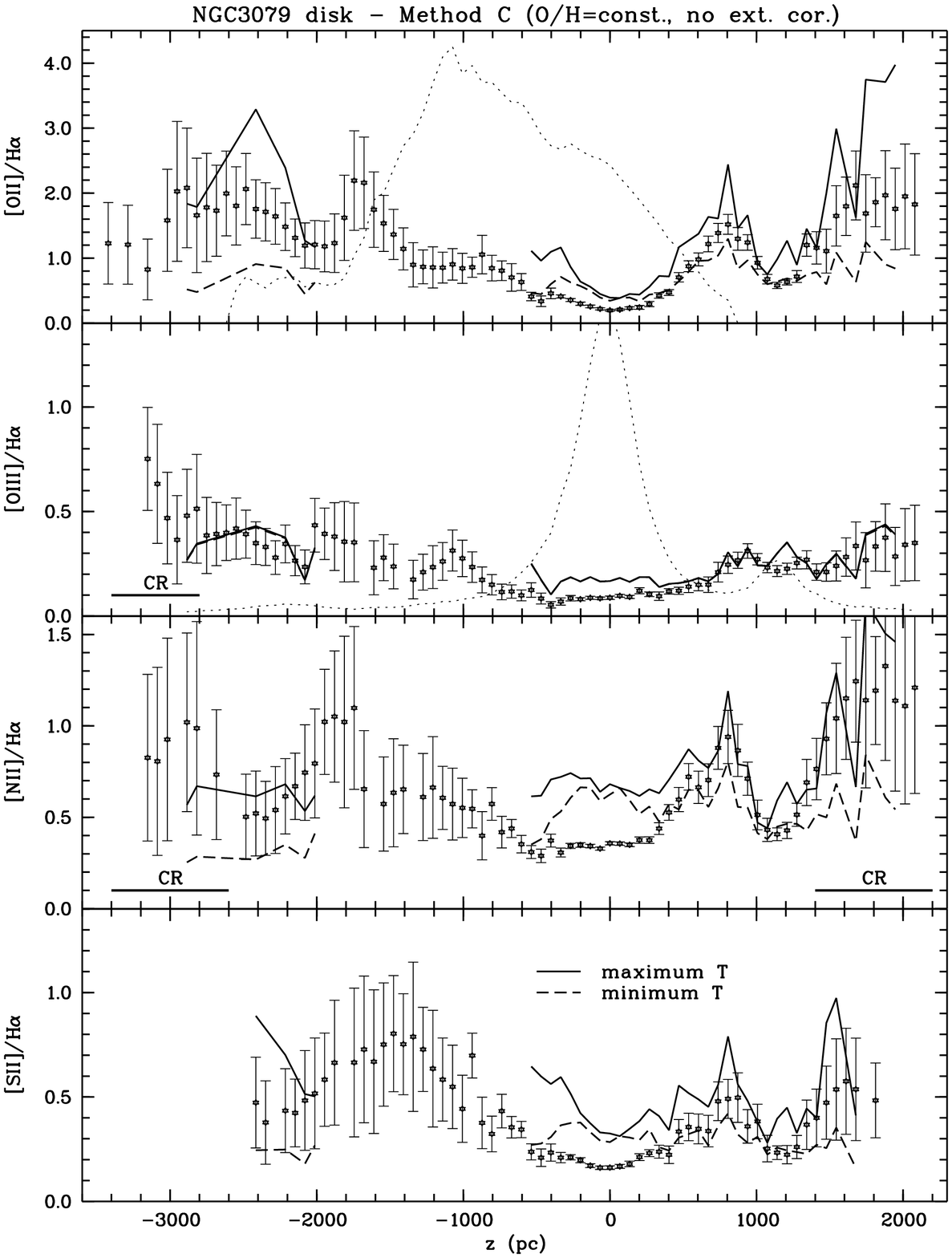}
\caption{}
\end{figure}
\begin{figure}
\figurenum{14d}
\plotone{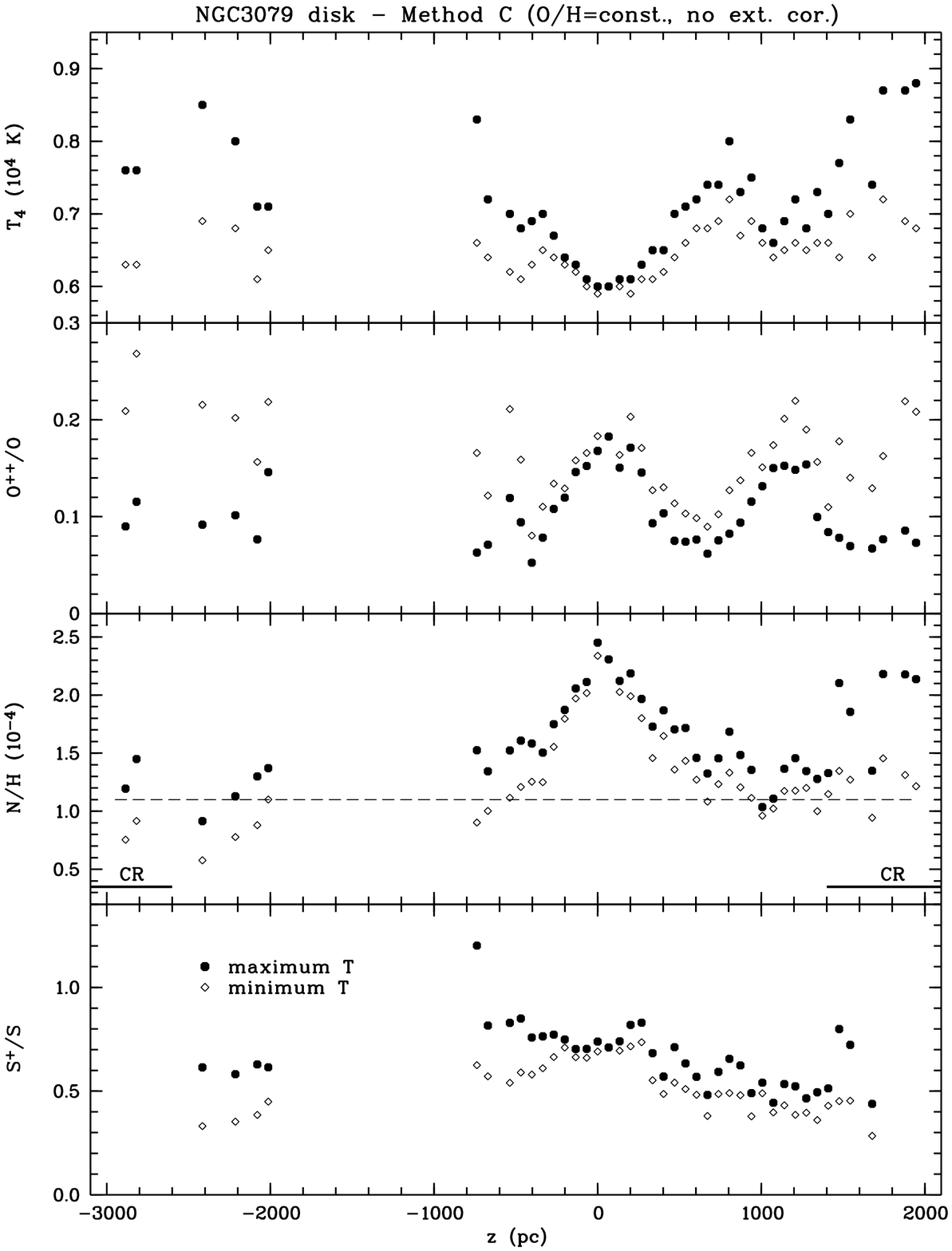}
\caption{}
\end{figure}
\begin{figure}
\plotone{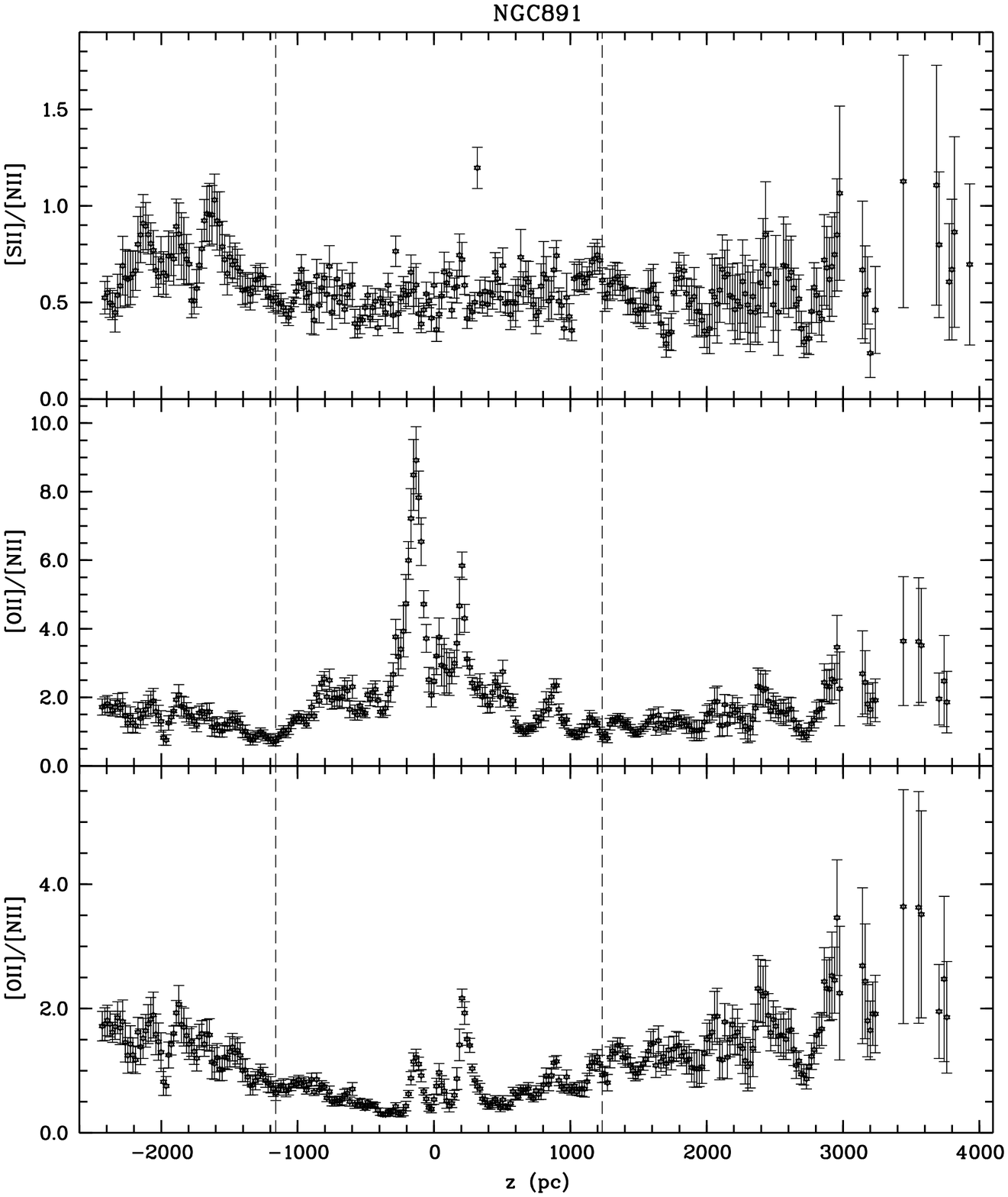}
\caption{[\protect\ion{S}{2}]/[\protect\ion{N}{2}] and [\protect\ion{O}{2}]/[\protect\ion{N}{2}] line ratios
for NGC\,891. {\em Top panel}: [\protect\ion{S}{2}]/[\protect\ion{N}{2}] (extinction corrected).
{\em Middle panel}: [\protect\ion{O}{2}]/[\protect\ion{N}{2}] (extinction corrected). {\em
Bottom panel}: [\protect\ion{O}{2}]/[\protect\ion{N}{2}] (without extinction correction). The
{\em dashed lines} show the range affected by the extinction correction.}
\end{figure}
\begin{figure}
\plotone{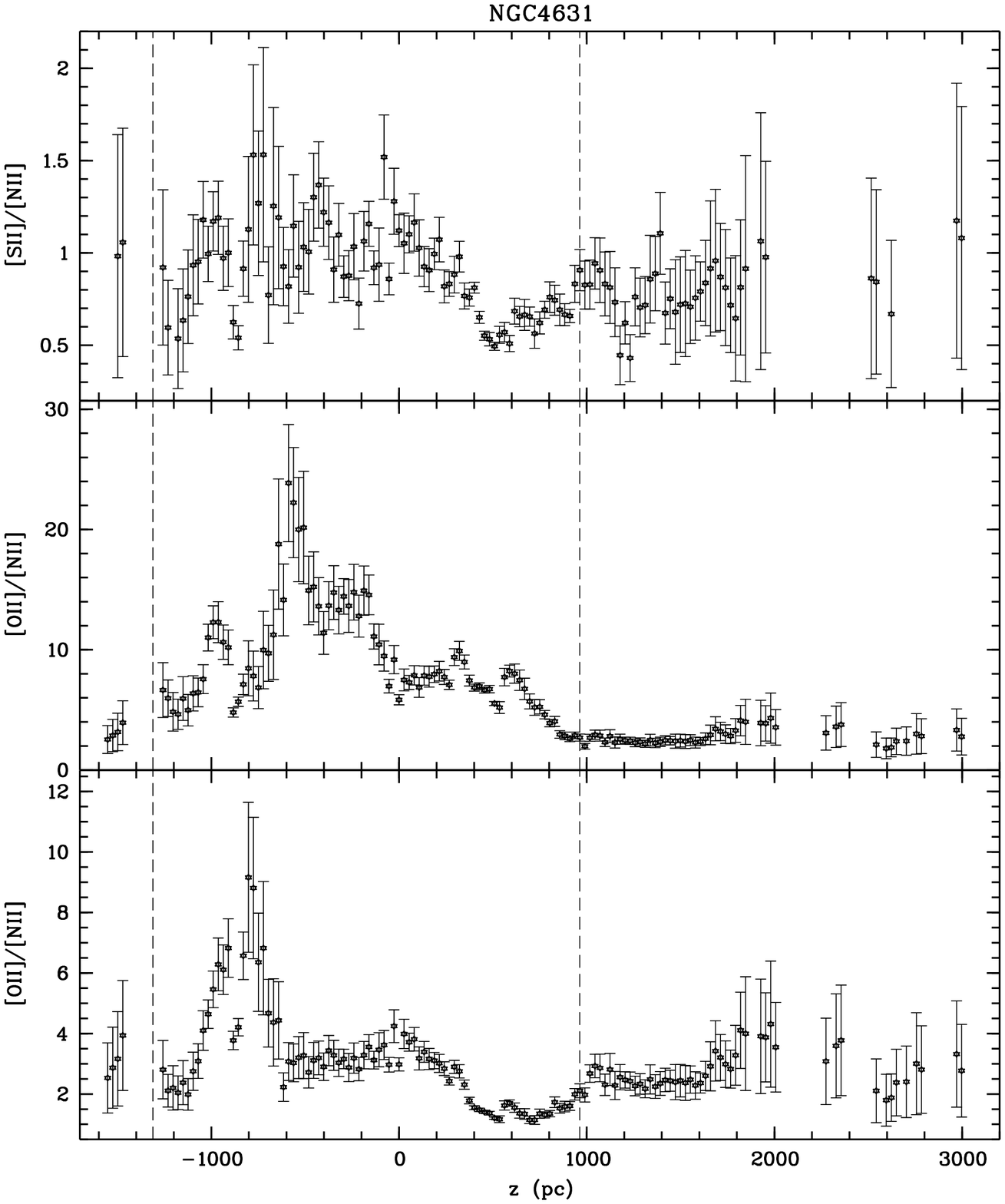}
\caption{Same as Fig. 15, but for NGC\,4631.}
\end{figure}
\begin{figure}
\plotone{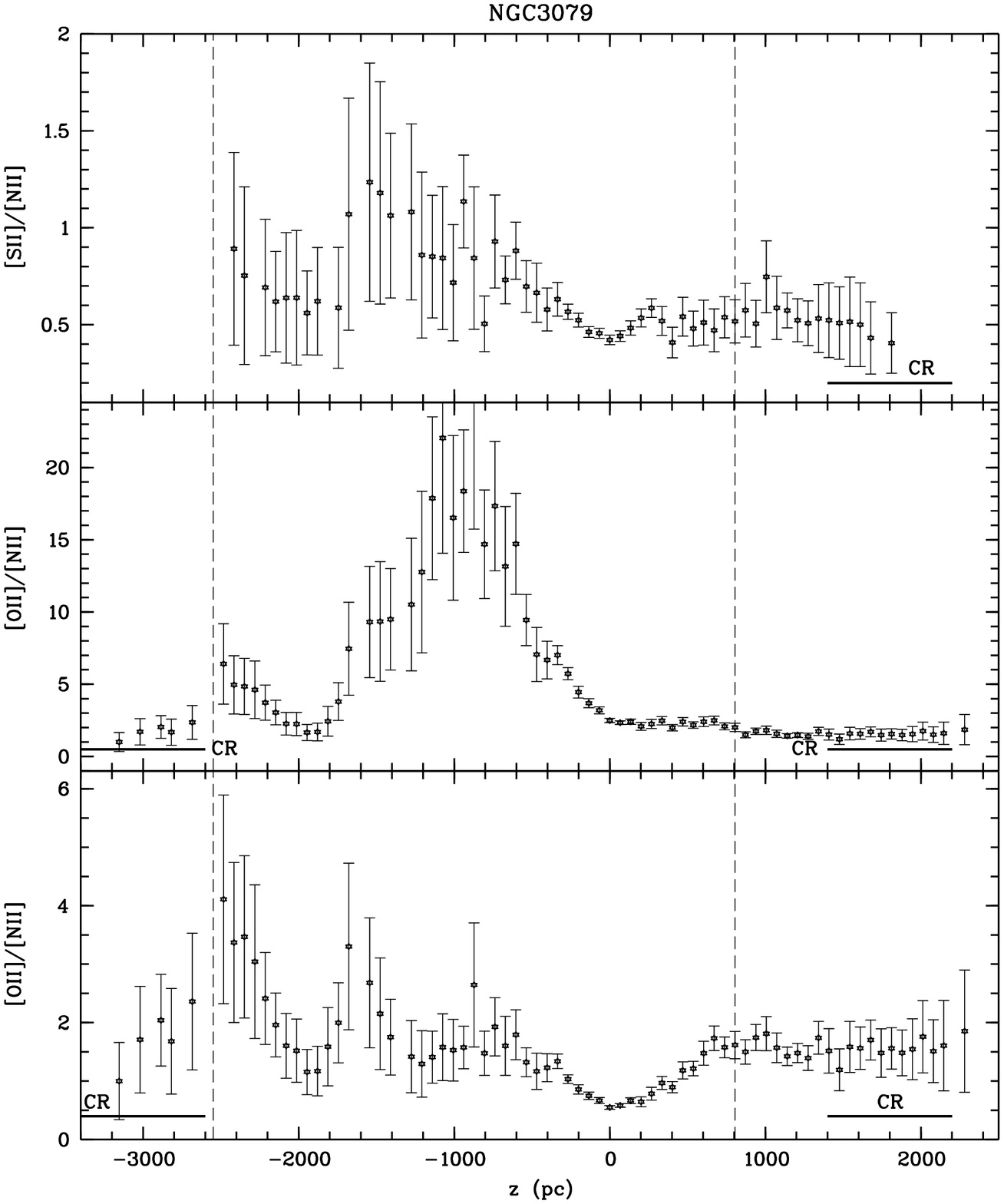}
\caption{Same as Fig. 15, but for NGC\,3079. Data contaminated by cosmic ray
hits are marked with the bars labeled ``CR''.}
\end{figure}
\end{document}